\documentclass[electronics]{Definitions/mdpi}
\firstpage{1} 
\pubvolume{xx}
\issuenum{10}
\articlenumber{4}
\pubyear{2021}
\copyrightyear{2021}
\history{Received: date; Accepted: date; Published: date}
\pdfoutput=1
\usepackage{subcaption}
\Title{Broadband RF Phased Array Design with MEEP: Comparisons to Array Theory in Two and Three Dimensions}
\Author{Jordan C. Hanson}
\AuthorNames{Jordan C. Hanson}
\address[1]{%
\quad Whittier College, Whittier, CA, USA; jhanson2@whittier.edu}

\abstract{Phased array radar systems have a wide variety of applications in engineering and physics research.  Phased array design usually requires numerical modeling with expensive commercial computational packages.  Using the open-source MIT Electrogmagnetic Equation Propagation (MEEP) package, a set of phased array designs is presented.  Specifically, one and two-dimensional arrays of Yagi-Uda and horn antennas were modeled in the bandwidth [0.1 - 5] GHz, and compared to theoretical expectations in the far-field.  Precise matches between MEEP simulation and radiation pattern predictions at different frequencies and beam angles are demonstrated.  Given that the computations match the theory, the effect of embedding a phased array within a medium of varying index of refraction is then computed.  Understanding the effect of varying index on phased arrays is critical for proposed ultra-high energy neutrino observatories which rely on phased array detectors embedded in natural ice.  Future work will develop the phased array concepts with parallel MEEP, in order to increase the detail, complexity, and speed of the computations.}


\begin{document}

\section{Introduction}
\label{sec:intro}

Radio-frequency phased array antenna systems with design frequencies of order 0.1-10 GHz have applications in 5G mobile telecommunications, ground penetrating radar (GPR) systems, and scientific instrumentation \cite{10.1109/iceaa.2017.8065458,10.3997/2214-4609.201701121,10.1088/1475-7516/2016/02/005,10.1109/iwem.2014.6963645}.   In the one-dimensional case, a series of three-dimensional antenna elements are arranged in a line with fixed spacing \cite{avva}.  Common antenna designs like loops and dipoles can be used to limit the \textit{elements} to two dimensions.  In this special case, phased array radiation may be modeled in two spatial dimensions plus time.  In the two-dimensional case, a series of three-dimensional antenna elements are arranged in a two-dimensional pattern, often a grid with fixed element spacing in both dimensions.  The elements may be strictly two-dimensional, but there is still an increase in computational complexity and the radiation is calculated in three dimensions plus time.

Proprietary RF modeling packages like XFDTD and HFSS are often used to model the response of elements within phased arrays and the behavior of arrays \cite{hfss, 10.1109/tmtt.2019.2919838, 10.1109/apusncursinrsm.2017.8072977,10.1109/aps.2006.1710856}.  The XFDTD package, for example, relies on the finite difference time domain (FDTD) method.  The FDTD approach is a computational electromagnetics (CEM) technique in which spacetime and Maxwell's equations are broken into discrete form.  One variant of the FDTD method is the conformal FDTD method (CFDTD), recently used to study phased array concepts on a large scale \cite{10.1109/aps.2006.1710856}.  The NEC2 and NEC4 family of codes relies on the method-of-moments (MoM) approach \cite{10.1109/aps.2004.1331976}.  Aside from the cost, a drawback of proprietary modeling software can be a lack of fine control over each individual object in the simulation.  Because Maxwell's equations are scale-invariant, in principle open-source FDTD codes designed for optical regimes could be re-purposd for RF design workflows.  One such open-source package is the MIT Electromagnetic Equation Propagation (MEEP) package \cite{10.1016/j.cpc.2009.11.008}.

A recent review \cite{10.3390/electronics8121506} covered how open radio design software like openEMS \cite{10.1002/jnm.1875}, gprMax \cite{10.1016/j.cpc.2016.08.020}, and the NEC2 family of codes \cite{10.1109/aps.2004.1331976} facilitate design workflows.  In this work, the radiation patterns of one-dimensional and two-dimensional phased array designs are simulated with the MEEP package.  MEEP takes advantage of the scale-invariance of Maxwell's equations.  Common MEEP applications are found in optical wavelength $\mu$m-scale designs, but scale-invariance allows the user to treat designs as cm-scale RF elements (see Appendix for details).  Although MEEP has been used to optimize antenna designs \cite{10.1016/j.jcde.2017.06.004}, this work appears to be the first to model entire phased arrays in MEEP with a variable index of refraction.  Two classes of phased array element are considered: Yagi-Uda and horn antennas.  The former is applied to single-frequency designs, while the latter is applied to broadband design.  Each element class is treated in both the one-dimensional and two-dimensional cases.  The phase-steering properties and radiation patterns of all designs are shown to match theoretical predictions.  The appropriate array theory is shown in Section \ref{sec:theory}, based on Chapter 1 of Reference \cite{mailloux3}.  Section \ref{sec:1d} contains comparisons between theory and simulation for one-dimensional cases, and Section \ref{sec:2d} contains the corresponding two-dimensional comparisons.  In Section \ref{sec:n}, the varying index of refraction is introduced. Results and future work are summarized in Section \ref{sec:summary}.

\begin{figure}
\centering
\includegraphics[width=12cm]{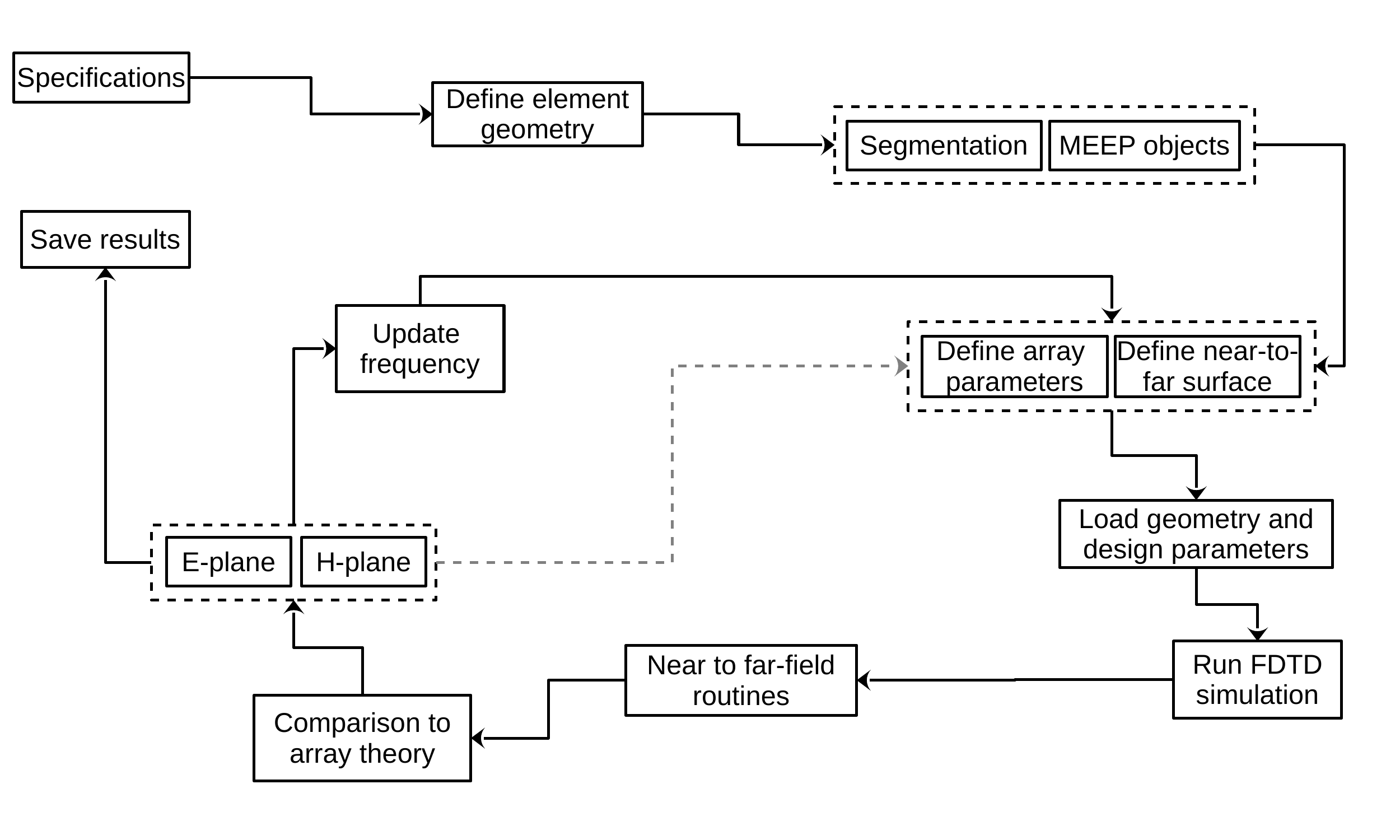}
\caption{\label{fig:flow} A detailed workflow for phased array design with MEEP.  See text for details.}
\end{figure}

A workflow using MEEP for phased array design is outlined in Figure \ref{fig:flow} based on Figure 1 from Reference \cite{10.3390/electronics8121506}.  Examples of decisions within the \textit{specifications} category are: single-frequency or broadband, desired directivity and beamwidth, side-lobe tolerance, and number of antenna elements.  These decisions lead to the choice of element type which must be implemented in MEEP.  Simple shapes like dipoles can be modeled with built-in MEEP objects.  Complex shapes like horns and dishes can be assembled from groups of objects.  Radiation sources and current functions must be defined.  For these studies, pure sinusoidal currents are passed to radiators which in turn radiate sinusoidal fields.  The dielectric constant and boundary conditions of the simulation volume and objects within the volume are defined in the next step.  The information is loaded into a simulation object and run for a number of time-steps.  Once complete, near-to-far field routines are called to produce the power at a set of angles.  The power versus angle is converted to normalized E and H-plane array radiation patterns and compared to theoretical models.  Given a match, the frequency is updated and the process is repeated.  If there is not a match, element separation and other array parameters are adjusted.

The workflow in Figure \ref{fig:flow} represents a non-parallelized approach.  Much development has gone into enhancing the speed, accuracy, and utility of the FDTD method.  First, MEEP itself may be run in parallel mode, providing a speed enhancement.  In a high-performance computing (HPC) environment, where each node has allocated memory (implying local RAM is not the limiting factor) running MEEP in parallel would speed up results.  There has also been CEM research devoted to enhancing the FDTD approach itself.  Decreasing memory usage and avoiding repetitive computations in favor of a more subtle approach is presented in \cite{10.1109/tmtt.2019.2919838}.  A three-dimensional implementation of FDTD algorithms on GPUs via CUDA has also been explored \cite{10.2528/pier13030606}.  The results of this work were obtained using the simplest version of MEEP: non-parallel with the python3 interface run in Jupyter notebooks on a laptop.  Therefore the results shown in Sections \ref{sec:1d} and \ref{sec:2d} could benefit from speed and memory enhancements in future studies.

\section{Phased Array Antenna Theory}
\label{sec:theory}

\begin{figure}
\centering
\includegraphics[width=0.7\textwidth]{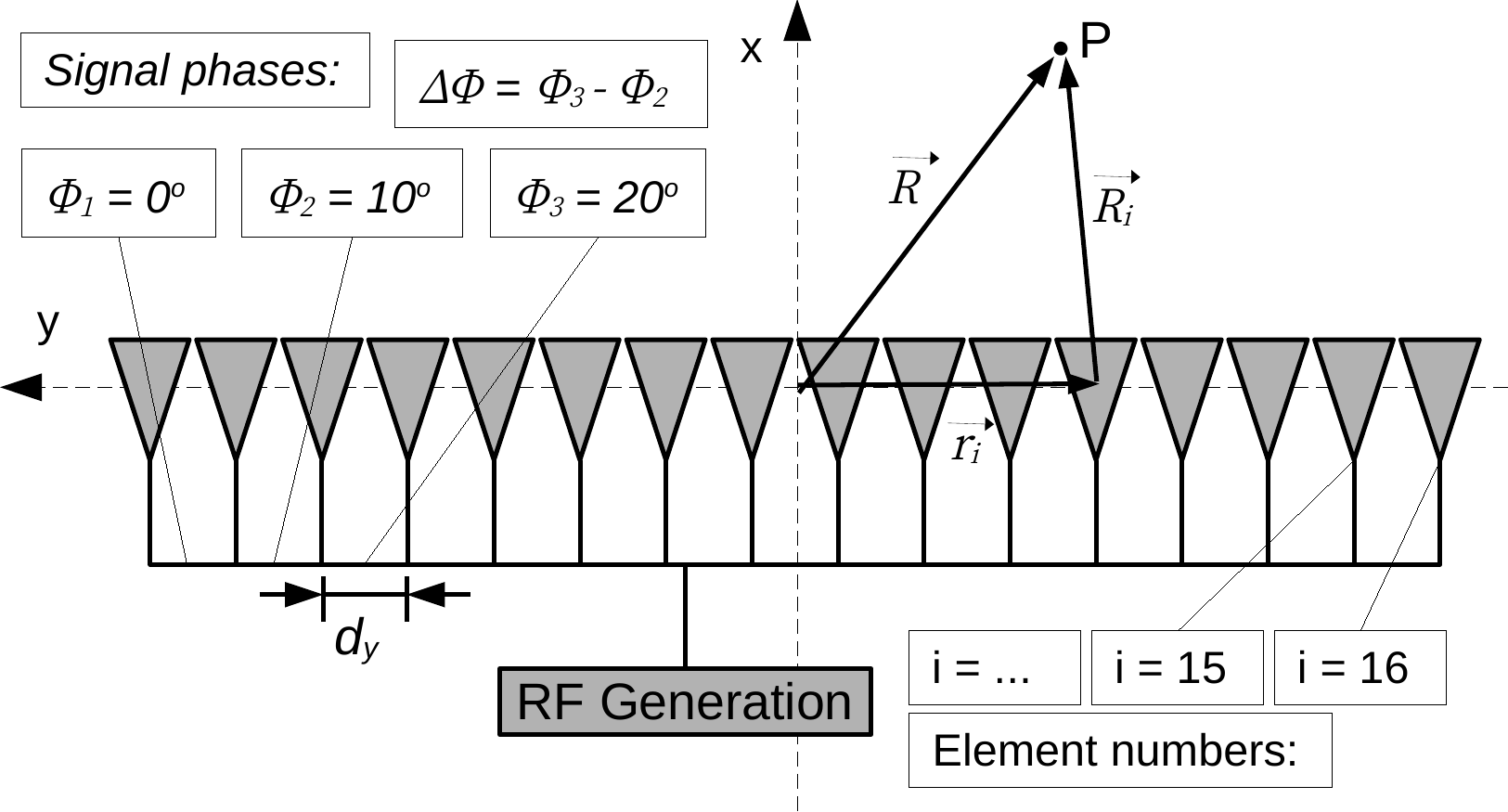}
\caption{\label{fig:array1} Definitions for the coordinate system, element label $i$, position vectors, and phase shift per antenna for a one-dimensional phased array of RF radiating elements. An example phase shift per antenna of $\Delta \Phi = \Phi_2 - \Phi_1 = \Phi_3 - \Phi_2 = \Phi_{i+1} - \Phi_i = 10^{\circ}$ value is shown.  Example position vectors for the 12th element are shown: $\vec{R} = \vec{r}_{12} + \vec{R}_{12}$.}
\end{figure}

The basic structure of a one-dimensional phased array of RF radiating elements is shown in Figure \ref{fig:array1}. Two important numerical constants that determine the beam angle $\Delta \phi$ of the array are the inter-element spacing $d_y$ and the phase shift per antenna $\Delta \Phi$.  Letting the subscript $i$ label each of the $N$ elements, the one-dimensional inter-element spacing in Figure \ref{fig:array1} is $d_y \hat{j} = \vec{r}_{i+1} - \vec{r}_i$, where $\vec{r}_i$ records the position of element $i$.  The phase shift per antenna is $\Delta \Phi = \Phi_{i+1} - \Phi_i$.  The relationship between $d_y$, $\Delta \Phi$, and $\Delta \phi$ is derived in Section \ref{sec:ba}.  The radiation pattern for a given $\Delta \phi$ is derived in Section \ref{sec:rp}.  For all coordinate systems, the azimuthal angle in the xy-plane is $\phi$, and the polar angle from the z-axis is $\theta$.

\subsection{Phase Steering and Beam Angle}
\label{sec:ba}

The beam angle $\Delta \phi$ of the array given $\Delta \Phi$ and $d_y$ will now be derived for the coordinate system in Figure \ref{fig:array1}.  First, the relevant far-field approximation will be described.  Second, it will be assumed that the elements all radiate at the same frequency $\omega$ and have the same vector radiation pattern $\vec{f}(\theta,\phi)$ that accounts for co-polarized and cross-polarized radiated power.  Third, the $\vec{E}$-field at point P will be treated as a sum of the $\vec{E}_i$ radiated from each element.  Fourth, the calculations will be restricted to the xy-plane and the relationship between the beam angle $\Delta \phi$ and array parameters will be obtained for a one-dimensional array.

According to Figure \ref{fig:array1}, the position of P can be written

\begin{equation}
\vec{R} = \vec{r}_i + \vec{R}_i
\end{equation}

Rearranging, the displacement between the i-th element and P is

\begin{equation}
\vec{R}_i = \vec{R} - \vec{r}_i
\end{equation}

The magnitude of the displacement is 

\begin{equation}
R_i = \sqrt{(\vec{R} - \vec{r}_i) \cdot (\vec{R} - \vec{r}_i) } = \left( R^2 - 2 \vec{R} \cdot \vec{r}_i + r_i^2 \right)^{1/2}
\end{equation}

Factoring an $R^2$, and neglecting the third term because it is small compared to the others,

\begin{equation}
R_i \approx R \left( 1 - \frac{2 \vec{R} \cdot \vec{r}_i}{R^2} \right)^{1/2}
\end{equation}

Expanding in a Taylor series to first order in $2 \vec{R} \cdot \vec{r}_i/R^2$, with $\hat{r} = \vec{R}/R$, yields

\begin{equation}
R_i \approx R \left( 1 - \frac{ \hat{r} \cdot \vec{r}_i}{R} \right)
\end{equation}

Distributing the $R$ gives the approximation:

\begin{equation}
R_i \approx R - \hat{r} \cdot \vec{r}_i
\label{eq:ithE}
\end{equation}

The electric field at $P$ due to the $i$-th element with individual radiation pattern $\vec{f}_i(\theta,\phi)$ is 

\begin{equation}
\vec{E}_i(R,\theta,\phi) = \frac{\vec{f}_i(\theta,\phi)\exp(-j k R_i)}{R_i}
\label{eq:ithE2}
\end{equation}

Substituting Equation \ref{eq:ithE} into Equation \ref{eq:ithE2}:

\begin{equation}
\vec{E}_i(R,\theta,\phi) = \vec{f}_i(\theta,\phi) \frac{\exp(-j k R)}{R} \exp(j k \vec{r}_i \cdot \hat{r})
\end{equation}

The element positions are written in Cartesian coordinates, while P is written in spherical coordinates using $u = \sin\theta \cos\phi$ and $v = \sin\theta \sin\phi$:

\begin{align}
\vec{r}_i &= \hat{x}x_i + \hat{y}y_i + \hat{z}z_i \\
\hat{r} &= \hat{x}u + \hat{y}v + \hat{z}\cos{\theta}
\end{align}

The total field $\vec{E}$ at P requires summing over elements.  The current delivered to the $i$-th element could have a potentially complex amplitude $a_i$.  The details of how the currents $a_i$ are converted to radiated $\vec{E}$-field are taken to be part of $\vec{f}(\theta,\phi)$.  The summation for $\vec{E}$ over elements is

\begin{equation}
\vec{E}(R,\theta,\phi) = \frac{\exp(-j k R)}{R} \sum_i a_i \vec{f}_i(\theta,\phi) \exp(j k \vec{r}_i \cdot \hat{r})
\end{equation}

For identical radiating elements: $\vec{f}_i = \vec{f}$:

\begin{equation}
\vec{E}(R,\theta,\phi) = \vec{f}(\theta,\phi) \frac{\exp(-j k R)}{R} \sum_i a_i \exp(j k \vec{r}_i \cdot \hat{r})
\end{equation}

Define the \textit{array-factor} $F(\theta,\phi) = \sum_i a_i \exp(j k \vec{r}_i \cdot \hat{r})$:

\begin{equation}
\vec{E}(R,\theta,\phi) = \vec{f}(\theta,\phi) \frac{\exp(-j k R)}{R} F(\theta,\phi)
\label{eq:totalE}
\end{equation}

Thus, if $F = 1$, then the $\vec{E}$-field is a plane wave, modified only by the elemental radiation pattern.  Complex amplitudes $a_i$ that cause a plane wave with wavevector pointed to $(\theta_0,\phi_0)$ are

\begin{equation}
a_i = |a_i| \exp(- j k \vec{r}_i \cdot \hat{r}_0) \label{eq:choice}
\end{equation}

The notation for beam angle $\Delta \phi = \phi - \phi_0$ will be introduced shortly.  For $\hat{r}_0$, $u_0$ and $v_0$ take the corresponding $\theta_0$ and $\phi_0$ for the angles: $\hat{r}_0 = \hat{x}u_0 + \hat{y}v_0 + \hat{z}\cos{\theta_0}$.  The angles $(\theta_0,\phi_0)$ correspond to the plane wave because the phases in the array factor in Equation \ref{eq:totalE} are cancelled by those in Equation \ref{eq:choice}, and the summation is over just the magnitudes $|a_i|$. For a \textit{linear array} in one-dimension, oriented along the y-axis as shown in Figure \ref{fig:array1}, $\theta_0 = \pi/2$ and $\vec{r}_i = i d_y \hat{y}$:

\begin{equation}
\vec{E}(R,\theta,\phi) = \vec{f}(\theta,\phi)\frac{\exp(-jkR)}{R}\sum_{i} a_i \exp\left( j k (i d_y v)\right)
\end{equation}

The summation is $F(\pi/2,\phi_0)$, $v = \sin(\phi)$ and $v_0 = \sin(\phi_0)$.  The weights $a_i$ may be arranged to produce a plane wave at $\phi_0$:

\begin{equation}
a_i = |a_i| \exp\left( -j k i d_y v_0 \right)
\end{equation}

The i-th phase of $\vec{E}$ in the array factor is

\begin{equation}
\Phi_i = k i d_y (\sin\phi - \sin\phi_0)
\end{equation}

The difference $\Delta \Phi = \Phi_{i+1} - \Phi_i$ for angles not far from the x-axis, $|\phi| < 1$ and $|\phi_0| < 1$, is

\begin{equation}
\Delta \Phi \approx d_y k (\phi - \phi_0) = 2\pi (d_y/\lambda) (\phi - \phi_0) = 2\pi (d_y/\lambda) \Delta \phi
\label{eq:lin}
\end{equation}

The \textit{beam angle} is $\Delta \phi = \phi - \phi_0$, the angular distance between a reference angle and the angle at which all contributions to $\vec{E}$ are in phase.  Equation \ref{eq:lin} reveals that the relationship between $\Delta \phi$ and $\Delta \Phi$ is linear, with slope $\lambda/(2\pi d_y)$.  In Section \ref{sec:1d}, the relationship between $\Delta \Phi$ and $\Delta \phi$ obtained from FDTD calculations via MEEP are shown to match precisely the theoretical prediction.  For two-dimensional grid arrays, the relationship ``factors,'' in that phase shift per element row and phase shift per element column govern $\Delta \phi$ and $\Delta \theta$ independently.  This theoretical prediction is matched precisely by the FDTD calculations shown in Section \ref{sec:2d} as well.

\subsection{Radiation Patterns and Beam Width}
\label{sec:rp}

The radiation pattern, or relative power $P$ emitted versus beam angle, is obtained from the array factor $F(\pi/2,\phi)$ summation.  Summation over the phased array with identical elements causes the vector element pattern $\vec{f}(\theta,\phi)$ and the common phase and amplitude factors $\exp(jkR)/R$ to cancel upon normalization.  The parameters that characterize the radiation patterns of arrays are $N$, the number of elements, and $d_y/\lambda$.  The magnitude of the complex current to each element is assumed to be the same, $|a_i| = a$.  Recall the array factor from Equation \ref{eq:totalE}, with $\theta = \pi/2$ and $a_i = a$:

\begin{equation}
F(\phi,\phi_0) = a \sum_{i} \exp\left( j k i d_y (v-v_0)\right) \label{eq:F}
\end{equation}

Let $\chi =  k d_y (v-v_0)$ so that $z = \exp\left( j \chi \right)$.  The sum is a geometric series from $i=1$ to $i=N$, the number of elements:

\begin{equation}
F(z) = a \sum_{i=1}^N z^i = a \left( \frac{1-z^N}{1 - z} \right)
\end{equation}

Using the Euler formula for $\sin(\chi)$, the array factor simplifies to

\begin{equation}
F(\chi) = -a \exp(j(N-1)\chi/2) \left( \frac{\sin(N\chi/2)}{\sin(\chi/2)}\right)
\end{equation}

The radiation pattern is proportional to power, so it is prudent to take the magnitude of $F(\phi)$:

\begin{equation}
|F(\chi)| = a \left( \frac{\sin(N\chi/2)}{\sin(\chi/2)}\right)
\end{equation}

The normalized radiation pattern will be $(F/F_{max})^2$, so it is necessary to find $F_{max}$:

\begin{equation}
\lim_{\chi \to 0} |F(\chi)| = a \lim_{\chi \to 0} \left( \frac{\sin(N\chi/2)}{\sin(\chi/2)}\right) = aN
\end{equation}

So $|F(\chi)|/F_{max}$ is

\begin{equation}
\frac{F(\chi)}{F_{max}} = \frac{\sin(N\chi/2)}{N \sin(\chi/2)}
\end{equation}

Finally, with $\chi = k d_y (v-v_0)$, $v = \sin(\phi)$, and $v_0 = \sin(\phi_0)$, the radiation pattern $P$ is $\left|F(\chi)/F_{max}\right|^2$:

\begin{equation}
\boxed{
P(\phi) = \left(\frac{\sin(\pi N (d_y/\lambda) (\sin(\phi) - \sin(\phi_0)))}{N \sin(\pi (d_y/\lambda) (\sin(\phi) - \sin(\phi_0)))}\right)^2
} \label{eq:radpatt}
\end{equation}

The -3 dB beamwidth is $0.886\lambda/L$, where $L = (N-1) d_y$.  In fact, Equation \ref{eq:F} is a function of $v-v_0$, so altering the $\Delta \Phi$ in the $a_i$ only rotates $P(\phi)$ in $\phi$-space, corresponding to a translation in \textit{v-space}.  The radiation pattern in Equation \ref{eq:radpatt} is shown to match precisely the main beam of FDTD calculations via MEEP for one-dimensional arrays in Section \ref{sec:1d}.  For two-dimensional grid arrays, the E and H plane radiation patterns ``factor,'' in that $P(\theta,\phi) = P(\theta)P(\phi)$.  In Section \ref{sec:2d}, precision matches for two-dimensional grid arrays are shown.

\subsection{Regarding Array Radiation Patterns}

Because one-dimensional and two-dimensional arrays are considered, some notes about radiation patterns are necessary.  First, all one-dimensional array radiation patterns correspond to the E-plane (the xy-plane).  The arrays are specified using elements situated in the xy-plane, and the array extends along the y-axis.  Radiators are linearly polarized such that the E-plane at some radius $r$ is $(r\cos(\phi),r\sin(\phi),0)$.  The H-plane at $r$ would be $(r\sin(\theta),0,r\cos(\theta))$, but this data is not relevant for a one-dimensional array.  Second, the MEEP python routine \verb+get_farfield+ is evaluated at a radius $r \gg L$, the length of the array, to obtain the far-fields $\vec{E}$ and $\vec{H}$.  Notice that not all open-source FDTD codes offer near-field to far-field transition modeling \cite{10.3390/electronics8121506}.

All two-dimensional phased array elements presented in Section \ref{sec:2d} are arrayed in the yz-plane, and the E and H-planes have the same definitions as the one-dimensional case.  However, the H-plane results have been shifted so that the main beam occurs at $\theta = 0$ degrees, rather than the expected 90 degrees.  This is purely for visual comparison to Equation \ref{eq:radpatt} cast as $P(\theta)$ with $\theta_0 = 0$, and does not mean the phased array is radiating orthogonally to broadside.  Equation \ref{eq:radpatt} is matched to E and H-plane two-dimensional patterns, and both are normalized to 0 dB at peak power.

\section{Phased Array Designs in One Dimension: Two-dimensional Fields}
\label{sec:1d}

Two antenna designs were considered in modeling one-dimensional phased arrays: Yagi-Uda and horn, corresponding to narrowband and wideband applications, respectively.  The two designs are depicted in Figure \ref{fig:ant1} with associated parameters described in Section \ref{sec:1d_ba} below.  The Yagi-Uda antennas have 6 elements with the same radius, oriented in the xy-plane: one reflector, one radiatior, three directors and a connecting boom.  The current $a_i$ is connected only to the radiator.  The horn antennas have three structures: the box containing the linearly polarized radiator, the radiator which is connected to $a_i$, and the curves of the horn.  An exponential function $y = f(x) = k_1 \exp(k_2 (x - a))$ describes the curves (see Figure \ref{fig:ant1}), and the origin is taken to be at the center of the back edge of the box.  The constants are $k_1 = a/2$ and $k_2 = (1/c) \ln(2d/a)$.  The curves are built from $n$ slices where $n = c/dx$.  All objects comprising the antenna elements have the same metallic conductivity, and the surrounding volume has an index of refraction $n=1$.  At the edge of the space is a layer 1 $\Delta x$ unit thick called the perfectly matched layer (PML) which cancels reflections.

\begin{figure}
\centering
\begin{subfigure}{0.45\textwidth}
\centering
\includegraphics[width=0.8\textwidth]{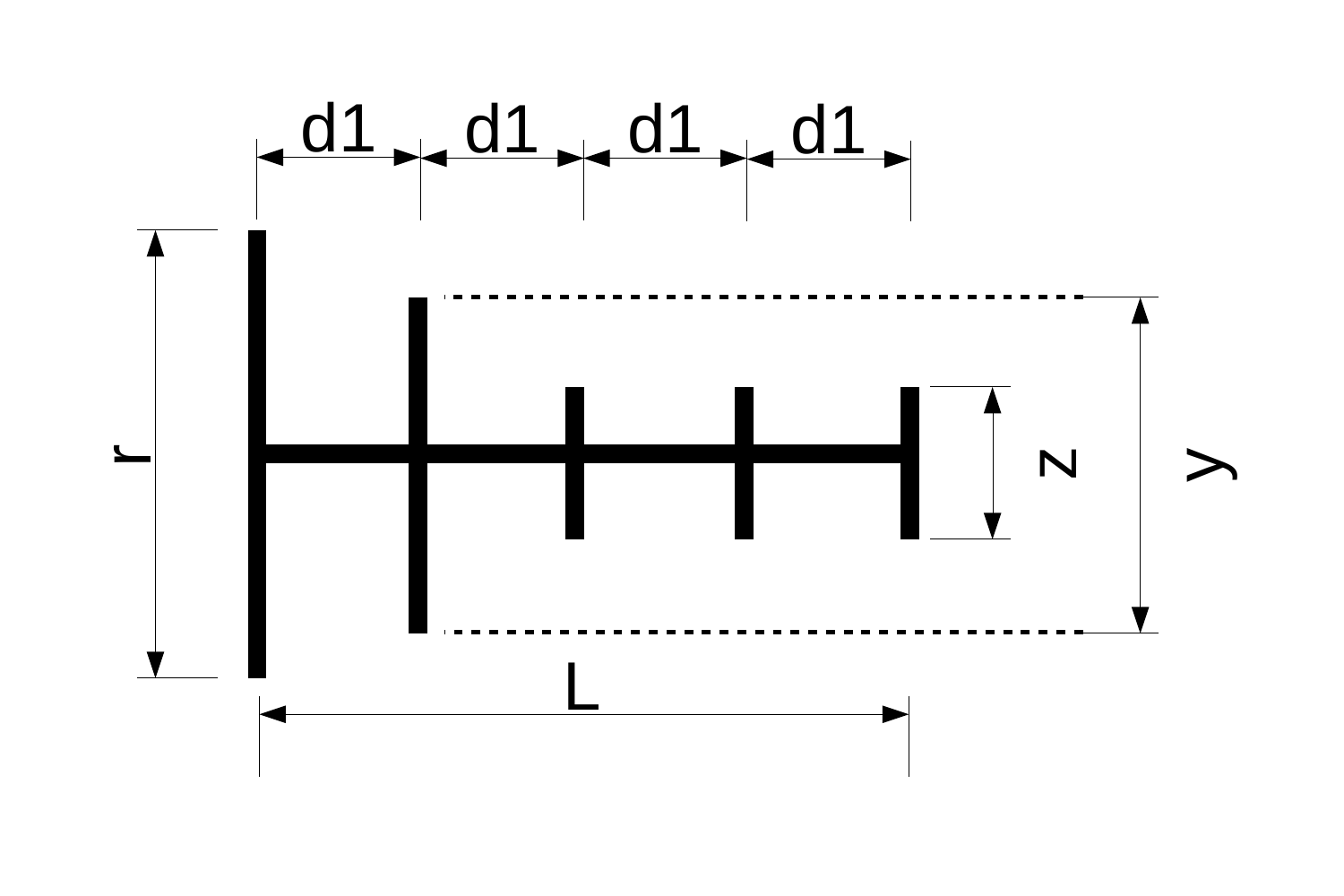}
\includegraphics[width=0.8\textwidth]{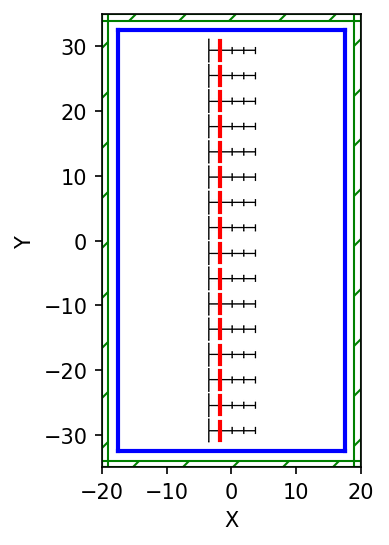}
\caption{The Yagi-Uda antenna, and the $N=16$ array.}
\end{subfigure}
\begin{subfigure}{0.45\textwidth}
\centering
\includegraphics[width=0.6\textwidth]{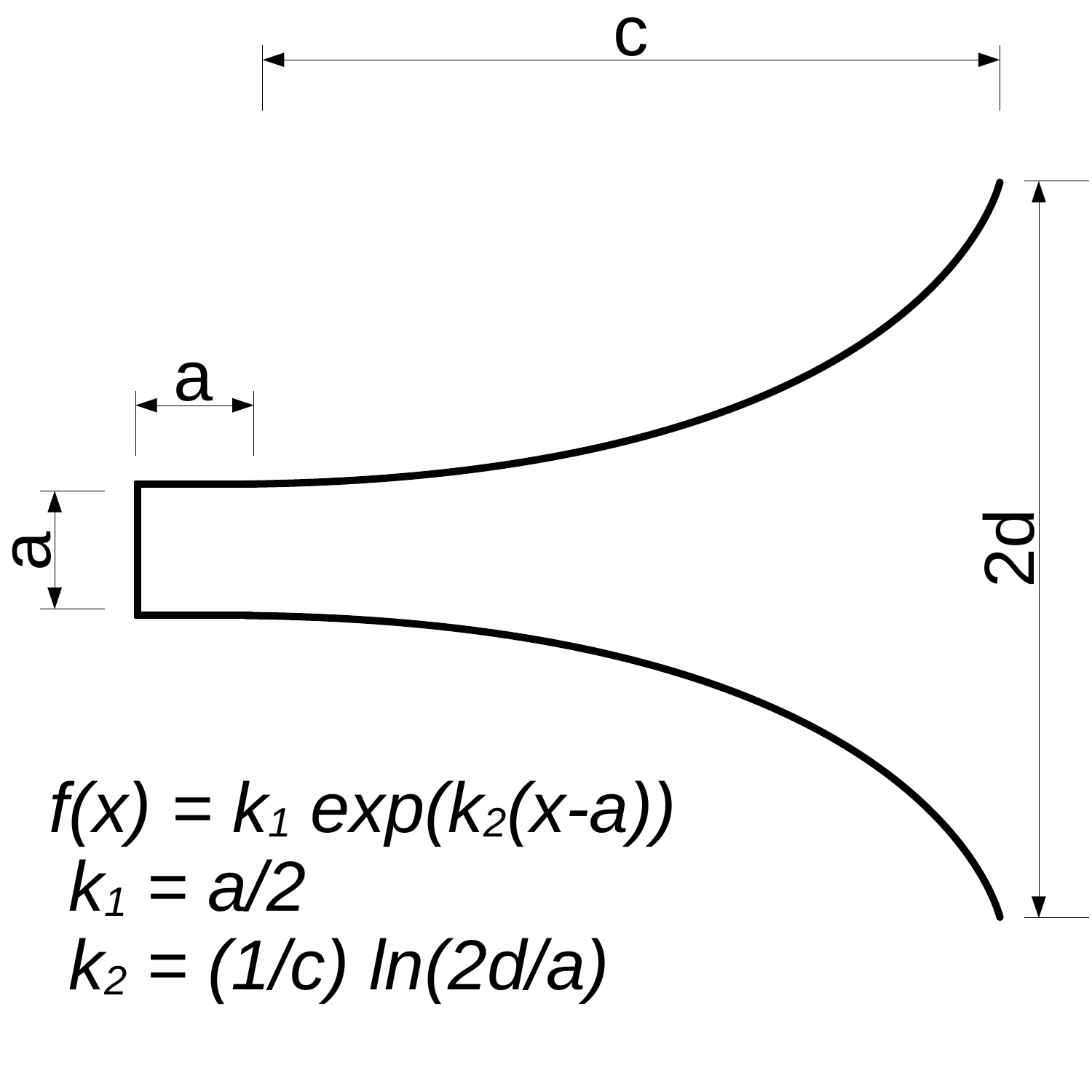}
\includegraphics[width=0.5\textwidth]{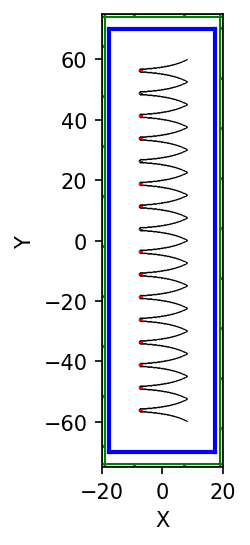}
\caption{The horn antenna, and the $N=16$ array.}
\end{subfigure} \vspace{0.5cm}
\caption{\label{fig:ant1} The two-dimensional antenna designs used in the one-dimensional phased array simulations.  (See Section \ref{sec:1d_ba} for details).  The black region has $\epsilon = \epsilon_0$, the green borders are perfectly matched layers (PML), and the blue surface is a MEEP \textit{Near2FarRegion} where flux is recorded for near-to-far projection.  The white lines represent metal structures, and the red lines represent the radiating elements.}
\end{figure}

\subsection{Phase Steering, Beam Angle, and Beamwidth}
\label{sec:1d_ba}

\begin{table}
\centering
\begin{tabular}{| c | c || c | c || c | c | c | c |}
\hline
\textbf{Yagi-Uda} & & \textbf{Horn} & & \textbf{Scan-loss, N=16 Horn array} & & & \\ \hline
\textit{Parameter} & \textit{Value} & \textit{Parameter} & \textit{Value} & Frequency (GHz) & $\Delta \Phi$ (degrees) & $d_y/\lambda$ & $SL_{dB}$ \\ \hline
$N$ & 8,16 & $N$ & 8,16 & 0.5 & 80 & 0.125 & -11.6 \\
$L$ & 7.20 & $a$ & 0.95 & 1.0 & 80 & 0.25 & -1.2 \\
$d_1$ & 1.80 & $c$ & 15.0 & 2.0 & 80 & 0.5 & -1.0 \\
$r$ & 3.75 & $d$ & 3.8 & 4.0 & 80 & 1.0 & -0.9 \\
$y$ & 2.81 & $dx$ & 0.1 & & & & \\
$z$ & 1.24 & $n=c/dx$ & 150 & & & & \\
$d_y$ & 3.92 & $d_y$ & $2d$ & & & & \\
resolution & 6 & resolution & 6 & & & & \\
\hline
\end{tabular}
\caption{\label{tab:runParam}  \textbf{Yagi-Uda}: The first and second columns contain the geometric parameters describing the antenna elements for the Yagi-Uda array.  \textbf{Horn}: The third and fourth columns contain those for the horn array.  \textbf{Scan-loss}: The fifth through eighth columns contain scan loss ($SL_{dB}$) data, reported for different frequencies and different $d_y/\lambda$ values for the $N = 16$ horn array.}
\end{table}

\begin{figure}
\centering
\includegraphics[width=0.49\textwidth]{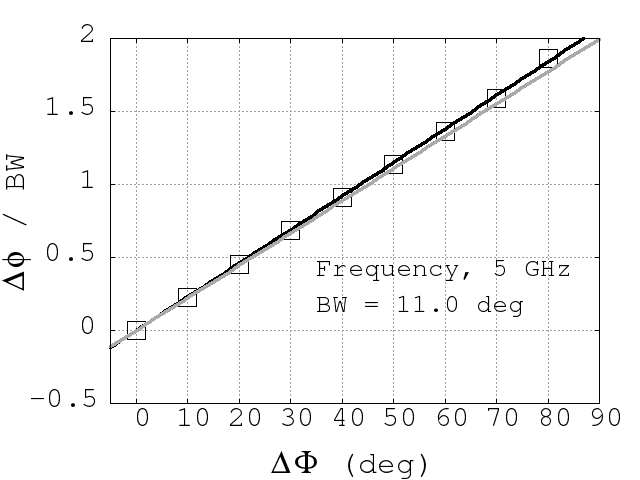}
\includegraphics[width=0.49\textwidth]{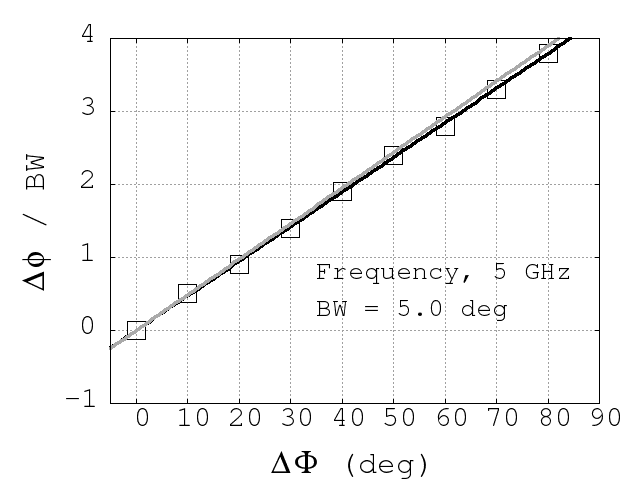} \\
\includegraphics[width=0.49\textwidth]{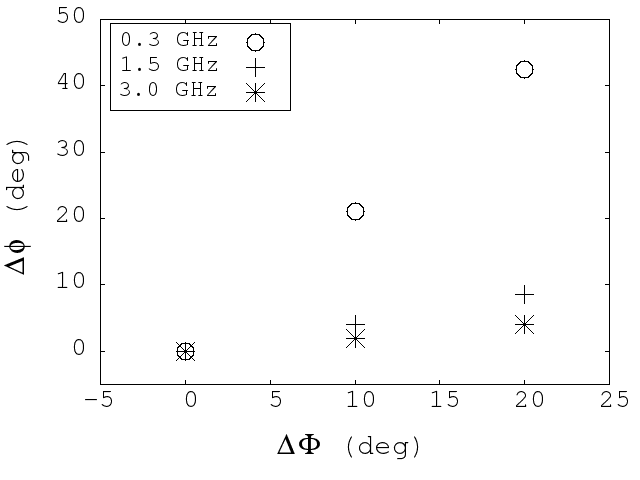}
\includegraphics[width=0.49\textwidth]{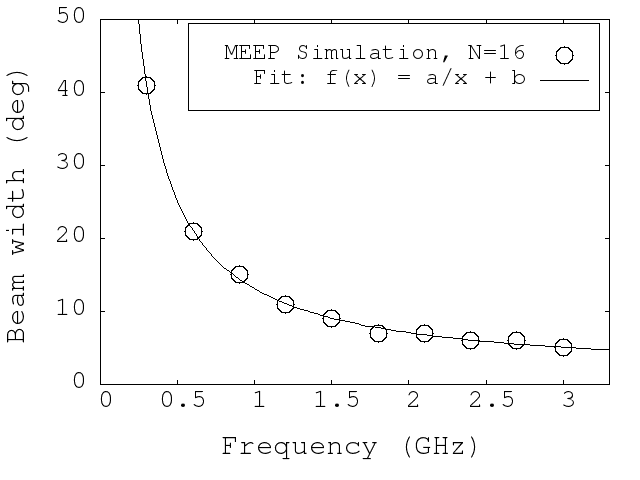}
\caption{\label{fig:dphase1} (Top left) The beam angle $\Delta \phi$ divided by the beam width $BW$ for the $N = 8$ one-dimensional Yagi array versus $\Delta \Phi$, the phase shift per element. (Top right) The same results for the $N = 16$ array. (Bottom left) $\Delta \phi$ versus $\Delta \Phi$ for the $N=16$ version of the one-dimensional horn array, for several frequencies.  (Bottom right) The dependence of the beam width on frequency for the one-dimensional $N=16$ horn array.}
\end{figure}

As described in Section \ref{sec:theory}, the beam angle is controlled by the phase shift per antenna.  Simulation results were run with the parameters in Table \ref{tab:runParam} for the one-dimensional arrays.  The main results are shown in Figure \ref{fig:dphase1}.  A discussion about scan loss below references data from Table \ref{tab:runParam}.

The phase-steering results are shown in Figure \ref{fig:dphase1}. The y-axes of Figure \ref{fig:dphase1} (top left) and (top right) are the beam angles of the Yagi-Uda arrays, divided by the beam widths.  The x-axes for these graphs are the phase shifts per element.  The top left plot and top right plots corresond to $N = 8$ and $N = 16$, respectively.  For the $N = 16$ horn case (bottom left and right), the value of $d_y/\lambda = f d_y/c$ varies because the elements can radiate from $\approx 0.3 - 5.0$ GHz.  The black solid lines in the top left and top right graphs of Figure \ref{fig:dphase1} are linear fits to the Yagi-Uda data. The gray lines represent the function $f(x) = b x$, with $b = \lambda/(2\pi d_y)$.  For these models, $d_y = 3.92$ cm and $\lambda = 6$ cm (Table \ref{tab:runParam}).  The slopes match almost exactly, with slight errors arising from radiation pattern distortion at high beam angle $\Delta \phi$.  At such large $\Delta \Phi$ values, side lobes can shift the location of the main beam by $\mathcal{O}(1)$ degree by merging with the main beam.  In the $N = 8$ case the fitted slope is slightly higher, and in the $N = 16$ case the fitted slope is slightly lower.  In each model, the observed beamwidths are within 1\% of the value predicted by Equation \ref{eq:lin}.

For the broadband horn case in the bottom left of Figure \ref{fig:dphase1}, three frequency cases are shown: 0.3, 1.5, and 3.0 GHz.  The intercepts are all consistent with zero and the slopes scale correctly: dividing the frequency by a factor of 2 increases the slope by a factor of 2, and dividing by a factor of 10 increases it by a factor of 10.  Graphs like the top left and top right of Figure \ref{fig:dphase1} would be misleading for horn antennas since the beamwidth depends on frequency (bottom right).  The fit parameters for beam width were $a = 12.0\pm 0.1$ degree GHz, and $b = 1.1 \pm 0.2$ degrees.  For these two-dimensional antennas, the $1/f$-dependence is a good description of the beamwidth across the [0.3 - 5 GHz] bandwidth.  The constant term $b$ is only necessary since the array has finite length $L$.  The beamwidth ($BW$) scales inversely with array length $L$: $BW \approx 0.886 \lambda/L$, from Equation \ref{eq:radpatt}.

A discussion of scan loss is merited when analyzing normalized radiation patterns, which are shown below in Section \ref{sec:1d_rp}.  Scan loss may be quantified as the peak power at the given beam angle divided by the peak power at a beam angle of zero degrees.  In the form of an equation in decibels, scan loss becomes a subtraction:

\begin{equation}
SL_{dB} = P_{\phi} - P_{\phi = 0}
\end{equation}

The scan loss $SL_{dB}$ is shown for the $N=16$ one-dimensional horn array in Table \ref{tab:runParam} (right), as it varies with frequency and $d_y/\lambda$.  The conservative value $\Delta \Phi = 80$ degrees was chosen because it is associated with the largest beam angles that do not generate side lobes larger than -15 dB.  Given the beam width of the $N=16$ design (5.04 degrees), this corresponds to a scan range of $\pm 20.16$ degrees.  The largest beam angles tend to have the largest scan losses, so the numbers in Table \ref{tab:runParam} should be the most conservative.  Scan losses of less than 1 dB are observed at high frequencies, but at $d_y/\lambda$ values that begin to admit large side lobes (Section \ref{sec:1d_rp}).

\subsection{Radiation Patterns}
\label{sec:1d_rp}

\begin{figure}
\centering
\includegraphics[width=0.35\textwidth]{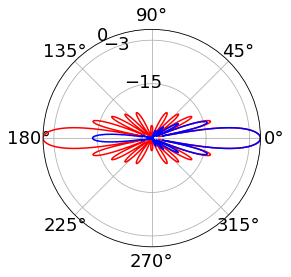}
\includegraphics[width=0.35\textwidth]{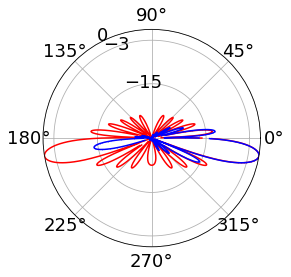}\\
\includegraphics[width=0.35\textwidth]{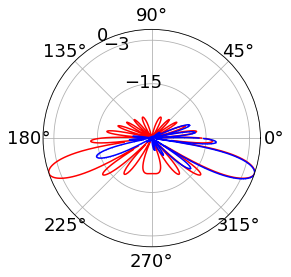}
\includegraphics[width=0.35\textwidth]{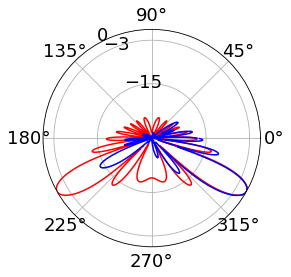}\\
\includegraphics[width=0.35\textwidth]{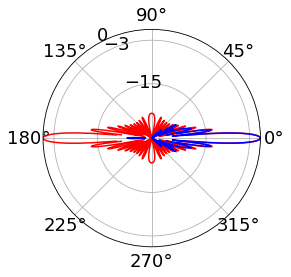}
\includegraphics[width=0.35\textwidth]{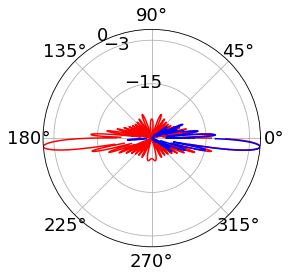}\\
\includegraphics[width=0.35\textwidth]{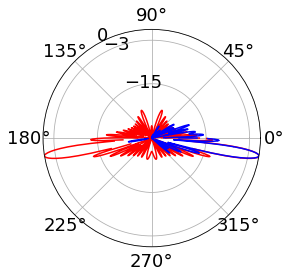}
\includegraphics[width=0.35\textwidth]{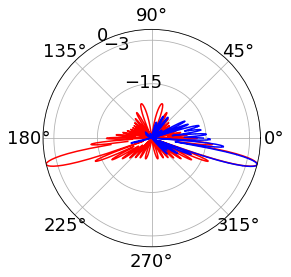}
\caption{\label{fig:1dyagiresults} \textbf{Yagi-Uda results, two-dimensional elements, one-dimensional array.}  (Top row) $f = 2.5$ GHz, and $\Delta \Phi = 0, 20$ degrees from left to right.  (Second row) $f = 2.5$ GHz, and $\Delta \Phi = 40, 60$ degrees from left to right.  (Third row) $f = 5.0$ GHz, and $\Delta \Phi = 0, 20$ degrees from left to right.  (Bottom row) $f = 5.0$ GHz, and $\Delta \Phi = 40, 60$ degrees from left to right.  The radial units are dB, and the angular units are degrees.}
\end{figure}

\begin{figure}
\centering
\includegraphics[width=0.35\textwidth]{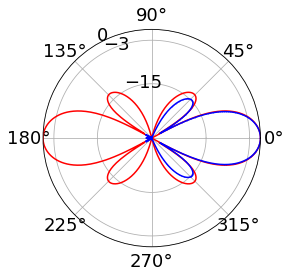}
\includegraphics[width=0.35\textwidth]{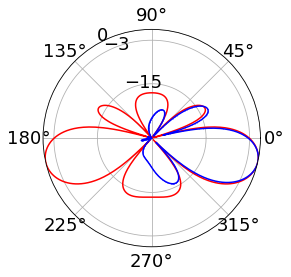}\\
\includegraphics[width=0.35\textwidth]{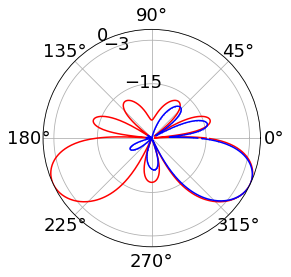}
\includegraphics[width=0.35\textwidth]{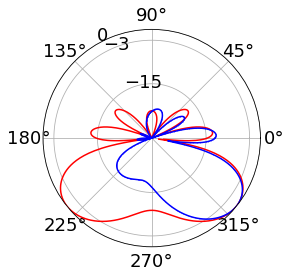}\\
\includegraphics[width=0.35\textwidth]{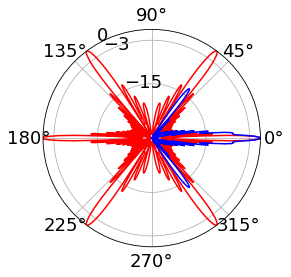}
\includegraphics[width=0.35\textwidth]{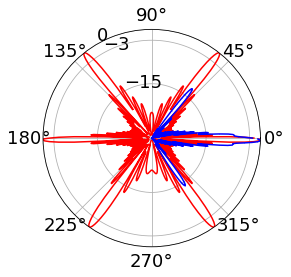}\\
\includegraphics[width=0.35\textwidth]{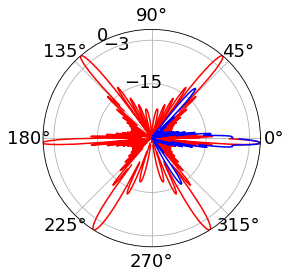}
\includegraphics[width=0.35\textwidth]{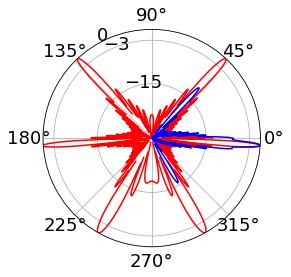}
\caption{\label{fig:1dhornresults} \textbf{Horn results, two-dimensional elements, one-dimensional array.}  (Top row) $f = 0.5$ GHz, and $\Delta \Phi = 0, 10$ degrees from left to right.  (Second row) $f = 0.5$ GHz, and $\Delta \Phi = 20, 30$ degrees from left to right.  (Third row) $f = 5.0$ GHz, and $\Delta \Phi = 0, 10$ degrees from left to right.  (Bottom row) $f = 5.0$ GHz, and $\Delta \Phi = 20, 30$ degrees from left to right.}
\end{figure}

Radiation patterns in the E-plane from $N=16$ one-dimensional Yagi and horn arrays are shown in Figures \ref{fig:1dyagiresults} and \ref{fig:1dhornresults}, respectively.  As described above, the x-direction ($\Delta\phi=0$) corresponds to no phase shift per element ($\Delta \Phi = 0$).  The radiation patterns are normalized to the power at the beam angle, and are shown in blue.  The red curves represent Equation \ref{eq:radpatt} with the correct $N$-value and $d_y/\lambda$-value.  Equation \ref{eq:radpatt} is symmetric, with identical forward and backward lobes.  The front-to-back or FB ratio would be 1.0 or 0 dB for a row of ideal point sources.  Although there is no backplane in either simulated one-dimensional array, the FB ratios of $\leq -15$ dB are observed.

The Yagi-Uda results are shown for 2.5 and 5.0 GHz frequencies in Figure \ref{fig:1dyagiresults}, with $\Delta \Phi = 0, 20, 40,$ and $60$ degrees.  Though the radiating elements are 6 cm long, good agreement between simulation and Equation \ref{eq:radpatt} is observed at both 2.5 GHz and 5.0 GHz, including side-lobes.  The beamwidth is proportionally larger at 2.5 GHz relative to 5.0 GHz, and at 5.0 GHz, the theoretical -3 dB beam width of 5.0 degrees is achieved.  The amplitudes of all side-lobes are limited to $\approx -15$ dB, except at the highest beam angles where scan losses are experienced.  Finally, the effect of frequency on beam steering is evident.  The same $\Delta \Phi$ does not generate as large a $\Delta \phi$ at higher frequencies because the slope implied by Equation \ref{eq:lin} is proportional to $\lambda$.

The horn results are shown in Figure \ref{fig:1dhornresults} for 0.5 GHz and 5.0 GHz frequencies, corresponding to the lower and upper end of the bandwidth.  The phase shifts per element are $\Delta \Phi = 0, 10, 20,$ and $30$ degrees.  The angular range of $\Delta \Phi$ is restricted relative to the Yagi-Uda case.  Wideband systems experience a natural trade-off in angular range versus bandwidth.  A $d_y/\lambda$ value that is acceptably smaller than one at low frequencies can grow larger with increasing frequency, leading to interference patterns.  At $5.0$ GHz, the horns radiate at $\pm 45$ degrees from $\Delta \phi = 0$.  The prediction from Equation \ref{eq:radpatt} is that these side-lobes, or \textit{grating lobes}, are equal in relative power to the main beam.  The actual array limits them to $-15$ dB, but only if $|\Delta \Phi| < 35$ degrees.  For larger phase shifts per element, the opposite side-lobe grows above $-15$ dB.  If the beam is steered too far in the $-\hat{\phi}$-direction, the side-lobe on the $\hat{\phi}$ side grows, and vice versa.

The general features of the radiation pattern compare well to the theoretical prediction.  The $1/f$-dependence of the main beamwidth is evident in Figure \ref{fig:1dhornresults}.  Like the Yagi-Uda array, the minimum theoretical beamwidth is reached at the highest frequencies (Figure \ref{fig:dphase1} bottom right).  The mini-lobes that are partially merged with the main beam widen the beam, however, the beamwidth is calculated at angles corresponding to -3 dB relative power.  Since the mini-lobes are below -3 dB, the beamwidth calculation is unaffected.  The simulation also matches the location and width of side-lobes to the theoretical prediction across the bandwidth.  The six grating lobes at 5 GHz are a result of the pattern multiplication theorem, which states that the normalized radiation pattern is a product of the horn pattern and the pattern of an array of point sources.  At 5 GHz, this multiplication suppresses the horn element pattern in the multiplication.

\begin{figure}
\centering
\includegraphics[width=0.282\textwidth,angle=90]{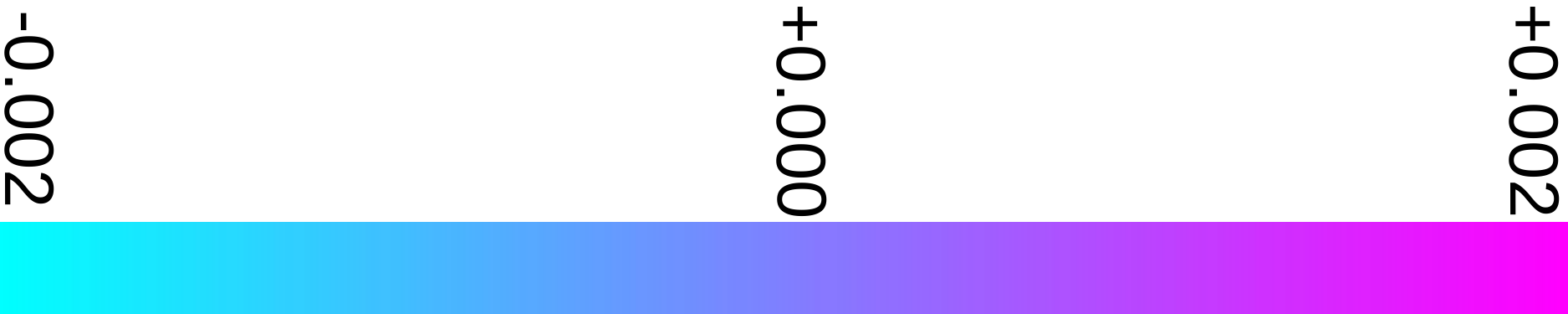}
\includegraphics[width=0.15\textwidth]{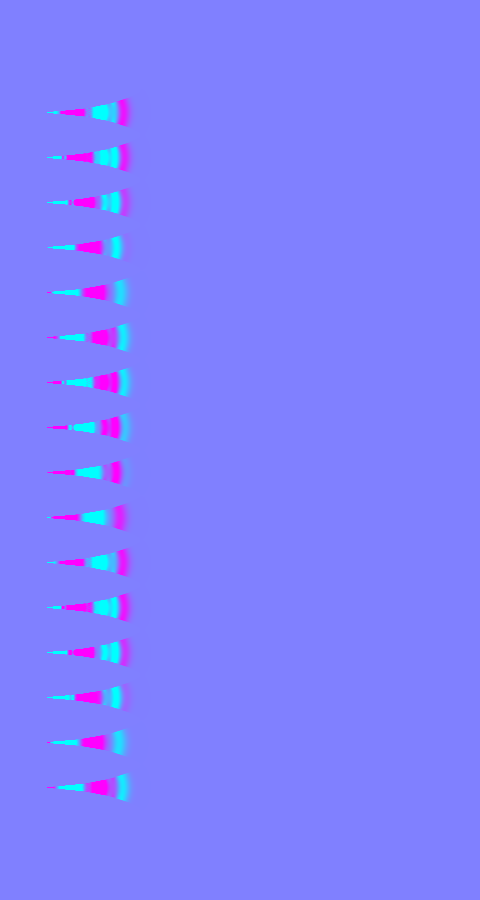}
\includegraphics[width=0.15\textwidth]{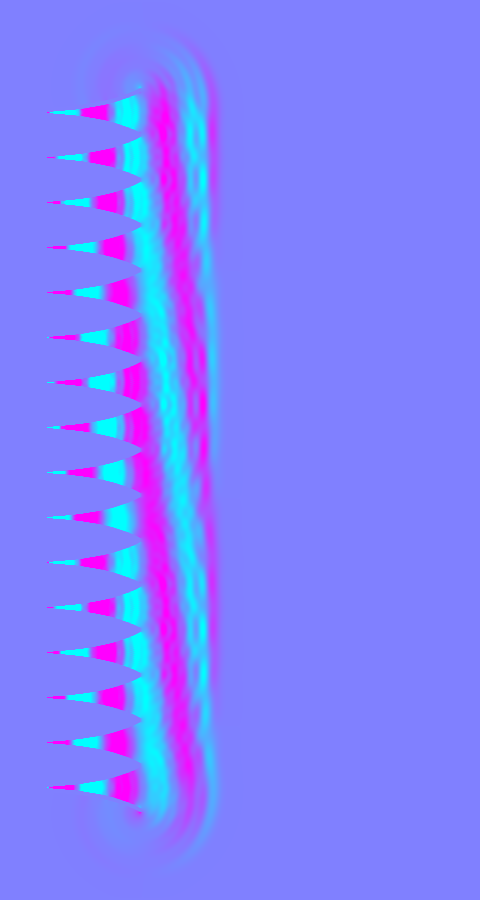}
\includegraphics[width=0.15\textwidth]{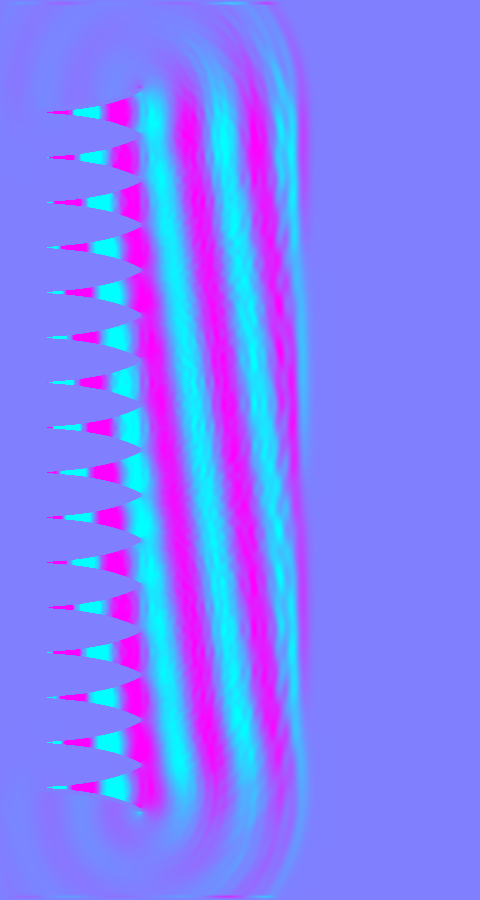}
\includegraphics[width=0.15\textwidth]{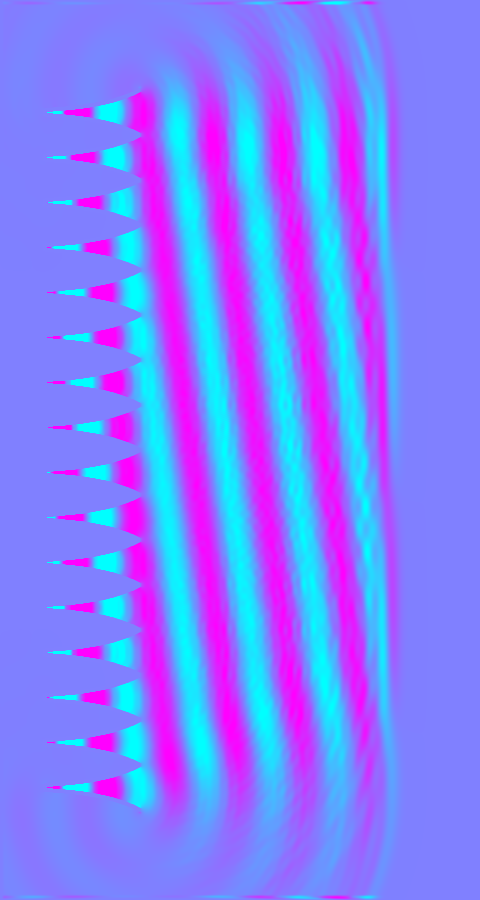}
\includegraphics[width=0.3\textwidth]{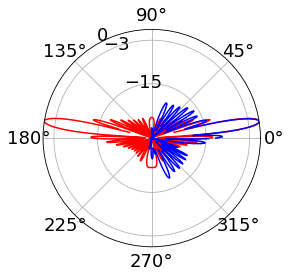}
\caption{\label{fig:1dhornresults2} (From left to right) The $N = 16$ horn one-dimensional linearly polarized electric field $|\vec{E}(x,y,t)|$ at $t = 0.5$ ns, $t = 1.0$ ns, $t = 1.5$ ns, and $t = 2.0$ ns.  The 2D area is $80 \times 150$ cm$^2$, with resolution of 6 pixels per distance unit ($480 \times 900$ pixels).  The frequency is 2.5 GHz, and the beam angle is 9 degrees from broadside. (Far right) The corresponding radiation pattern.}
\end{figure}

The field magnitude $|\vec{E}(x,y,t)|$ for the $N = 16$ horn array is shown in Figure \ref{fig:1dhornresults2} for $t = 0.5,1.0,1.5,2.0$ ns at a frequency of $2.5$ GHz, along with the radiation pattern (far right).  The original amplitude of the radiation source within the horns is $\pm 1$ units, and the color scale for radiated $|\vec{E}(x,y,t)|$ is $\pm 0.002$.  The $\Delta \phi$ is $9$ degrees, with $\Delta \Phi = -35$ degrees, and the beamwidth is $5.5 \pm 0.5$ degrees.  The area is $480 \times 900$ pixels describing $80 \times 150$ cm$^2$ with a resolution of 6 pixels per $\Delta x$.  The dimensions of the box (Tab. \ref{tab:runParam}: $a = 0.95$ cm) are smaller than $\lambda = 12$ cm.  As the radiation escapes to free space, the wavefront forms several $\lambda$ in front of the horns.  Higher-frequency modes with $f\gg 2.5$ GHz are observed at the wavefront that correspond to start-up effects at $t=0$ ns.  With 72 pixels per wavelength, minute features of $|\vec{E}(x,y,t)|$ can be interpreted as physical rather than numerical.  These features can be eliminated with amplitude smoothing near $t=0$ ns, a feature available in the MEEP \verb+CustomSource+ class.  Amplitude smoothing, however, makes the location of the wavefront less precise.

The radiation pattern in Figure \ref{fig:1dhornresults2} matches the theoretical prediction for the main lobe and first few side lobes.  The origin of the side lobes is apparent from the $|\vec{E}(x,y,t)|$ images, where diffraction patterns at the edges of the array are visible.  At 0.5 ns, radiation in an element is confined to the horn.  By 1.0 ns, that radiation joins the waves from horns on either side.  However, horns at the end of the array have no partner on one side, and some radiation leaks outside the main lobe.  The side lobes can change if the total run time is not sufficient.  The \verb+get_farfield+ routine in MEEP requires \verb+Near2FarRegion+ surfaces that form the near-field box that collects flux information at the radiation frequency.  The parameters of the near-field box are set by the \verb+add_near2far+ routine.  The \verb+get_farfield+ routine performs a near-to-far field projection to the given radius ($r = 1000$ cm) where the field is computed for the radiation pattern.  The propagation code must be run for sufficient units of $\Delta t$ so that enough radiation can cross the near-field box.  The code is run for 6.67 ns to generate the radiation pattern in Figure \ref{fig:1dhornresults2}.  Thus, the side lobes are averaged over many radiation periods.

\section{Phased Array Designs in Two Dimensions: Three-dimensional Fields}
\label{sec:2d}

For \textit{two-dimensional} grids of radiating elements, the array-factor $F(u,v)$ \textit{factors}:

\begin{equation}
F(\theta,\phi) = F(u-u_0) F(v - v_0) \label{eq:2d}
\end{equation}

\begin{figure}
\centering
\includegraphics[width=0.6\textwidth]{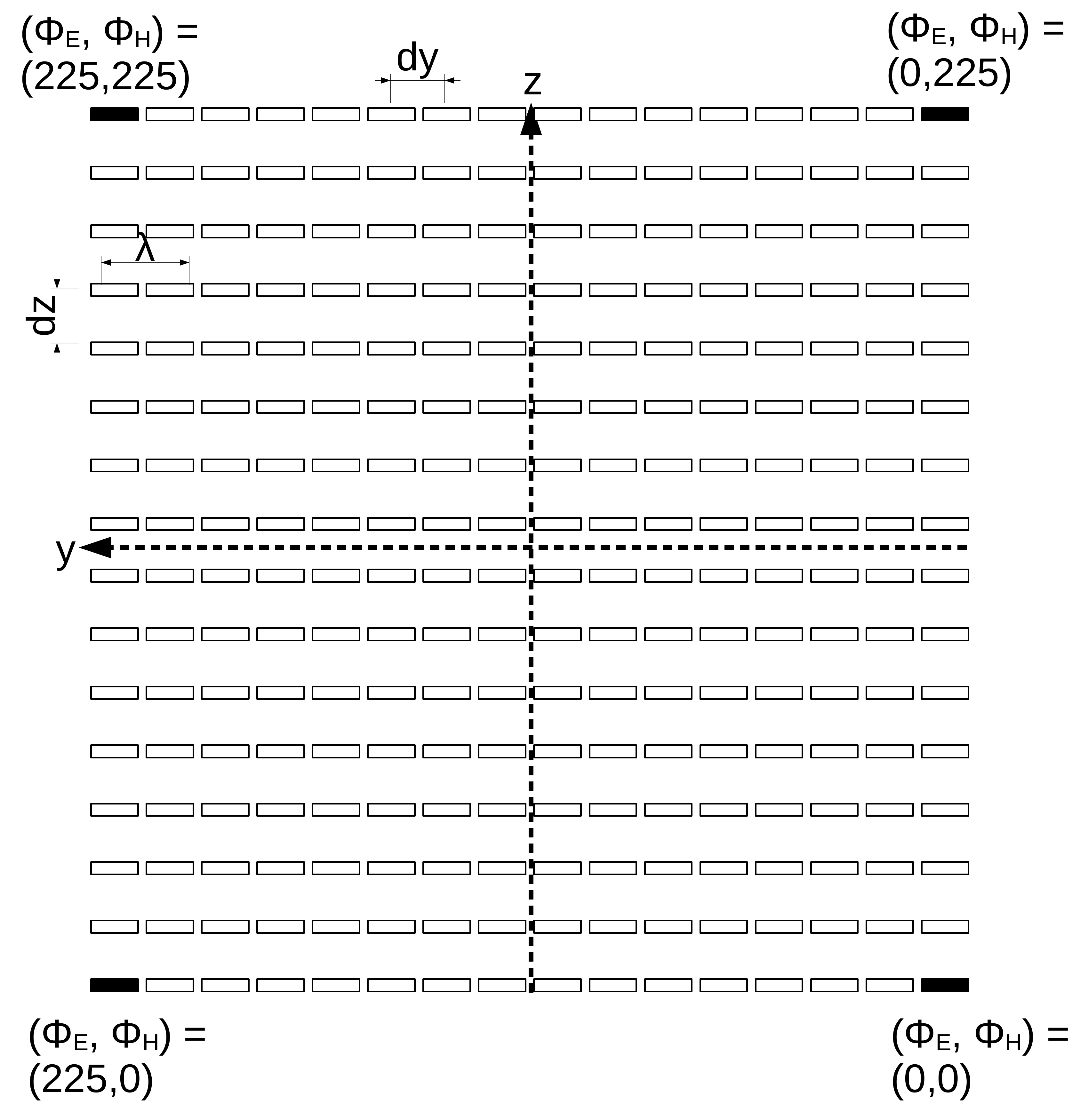}
\caption{\label{fig:schem2d} The two-dimensional $N \times N = 16 \times 16$ Yagi-Uda/horn y-polarized array layout.  The alignment with 3D Cartesian coordinates is depicted, along with the array spacing variables $d_y$ and $d_z$, the wavelength $\lambda$ (to scale), and the phases for each row and column if  $\Delta \Phi_E = \Delta \Phi_H = 15$ degrees.}
\end{figure}

The radiation pattern in Equation \ref{eq:radpatt} applies to the E and H plane separately.  The two-dimensional arrays modeled below are square $N \times N$ arrays, so beamwidths implied by Equation \ref{eq:radpatt} are equal for the E and H planes.  The complex phasing of Equation \ref{eq:2d} also indicates that $\Delta \phi_E \propto \Delta \Phi_E$, and $\Delta \phi_H \propto \Delta \Phi_H$, as shown in Equation \ref{eq:lin} for the one-dimensional case. For the designs presented, the H-plane corresponds to the xz-plane, and to varying the phase in the z-direction (by array row).  The E-plane corresponds to the xy-plane, and to varying the phase in the y-direction (by array column).  In Figure \ref{fig:schem2d}, the basic shape of the two-dimensional array is shown in the yz-plane with $\Delta \Phi_E = 15$ degrees, and $\Delta \Phi_H = 15$ degrees.  Section \ref{sec:2d_ba} contains results along the lines of Section \ref{sec:1d_ba} but for two-dimensional Yagi and horn arrays, and Section \ref{sec:2d_rp} contains results along the lines of Section \ref{sec:1d_rp} but for two-dimensional Yagi and horn arrays.  As before, Table \ref{tab:runParam} contains the typical run parameters, with a few important exceptions.

The first exception is that for the two-dimensional case $N$ becomes $N \times N$.  However, squaring the number of antennas raises the memory requirements.  In order to stay within a 16 GB memory limit, the two-dimensional \textit{horn array} results had to be restricted to $N \times N = 8 \times 8$.  The 2D horn array still has over $\mathcal{O}(10^4)$ metal objects, compared to the $\mathcal{O}(10^3)$ objects for the $N \times N = 16 \times 16$ 2D Yagi-Uda. The typical memory consumption is listed in Table \ref{tab:runParam2}, along with modified run paramters.  Further, the resolution parameter was restricted to 4.0 for the horns.  Restricting to 4 pixels per $\Delta x$ unit limits memory consumption, but then the box containing the radiator has too few pixels.  Enlarging the box allows the proper sized radiator to be fully contained.  A final object was added to reduce the FB ratio: a back-plane with parameters listed in Table \ref{tab:runParam2}.  One interesting modification is the doubling of the ratio of the box size ($a$) and the final horn width ($d$).  This had the effect of limiting the maximum frequency to $\approx 1$ GHz.  At 1 GHz, $d/\lambda \approx 0.5$.  A full optimization study on the horn parameters is warranted, though outside the present scope.

\begin{table}
\centering
\begin{tabular}{| c | c | c |}
\hline
\textbf{Horn} & & \\ \hline
\textit{Parameter} & \textit{Value} & \textit{System Information} \\ \hline
$N \times N$ & $8 \times 8$ & Memory Consumption \\
$a$ & 2.0 & 11.7 GB out of 15.5 GB \\
$c$ & 15.0 & \\
$d$ & 8.0 & CPU cores \\
$dx$ & 0.5 & Intel i7 1.80 GHz (8) \\
$n = c/dx$ & 30 & \\
$d_y$ & 16 & MEEP installation \\
resolution & 4 & Python3 interface (conda) \\
backplane location & $-2 a$ & \\
backplane thickness & 0.5 & \\
backplane dim. & 142 $\times$ 142 & \\
\hline
\end{tabular}
\caption{\label{tab:runParam2}  The parameters for the $N \times N = 8 \times 8$ horn array, modified from Table \ref{tab:runParam}.  The Yagi-Uda $N \times N = 16 \times 16$ array did not require modification.  The number of CPU cores was 4 in hardware, but was effectively 8 with hyperthreading.  The most memory-intensive simulation was the $8 \times 8$ horn array, which consumed 11.7 GB of memory out of 15.5 GB free.  The code was written with the Python3 interface to MEEP, installed with the conda package manager, and run in Jupyter notebooks.}
\end{table}

The horn elements radiate linearly polarized radiation in the y-direction, so the width in the z-direction does not follow the exponential functions but remains fixed at $a$.  Initial runs were performed with horn elements that simultaneously widened according to the exponential function defined in Section \ref{sec:1d}.  That design allowed reflections internal to the horns to distort the initial wavefront.  Holding the horn-width constant in the z-direction produces radiation patterns that match Equation \ref{eq:radpatt} because it follows the one-dimensional example of Section \ref{sec:1d}.  To obtain z-polarized wavefronts, all that is necessary is to rotate the array.  Practically, there are already examples of dually polarized RF band horns used in particle astrophysics \cite{10.1016/j.astropartphys.2009.05.003,undefined}, meaning that if this design were created with such elements, no rotation would be necessary.

\subsection{Phase Steering, Beam Angle, and Beamwidth}
\label{sec:2d_ba}

\begin{figure}
\centering
\includegraphics[width=0.49\textwidth]{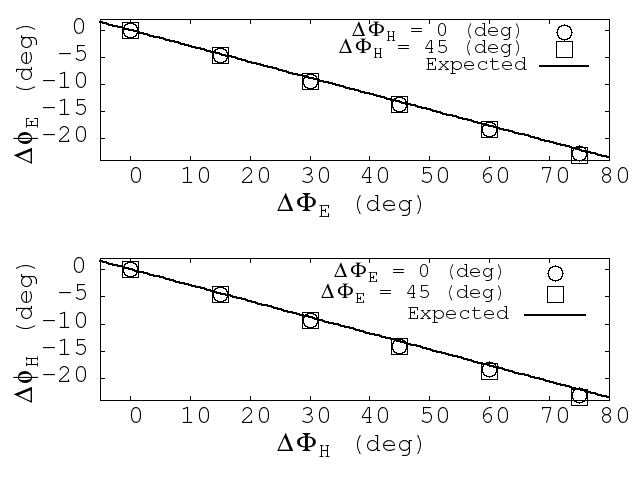}
\includegraphics[width=0.49\textwidth]{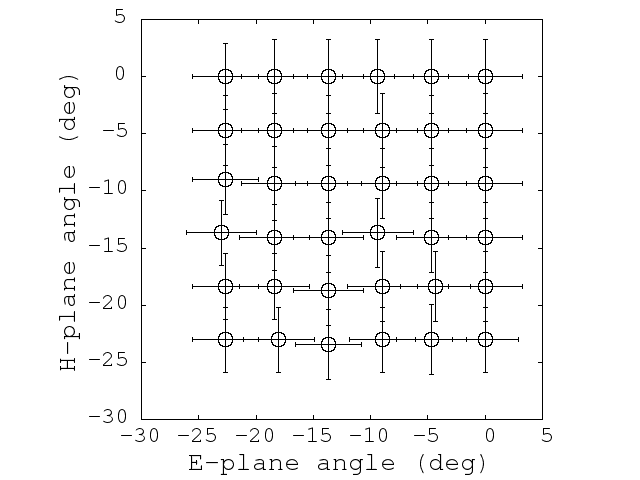}
\caption{\label{fig:phase2d} (Left) The beam angles $\Delta \phi_E$ and $\Delta \phi_H$ versus the phase shifts for the $N \times N = 16 \times 16$ Yagi-Uda array at 5 GHz.  The black lines represent the theoretical prediction of a linear dependence with slope $\lambda /(2\pi d_y)$ or $\lambda /(2\pi d_y)$, ($d_y = d_z$).  In each graph, the circles and squares correspond to two different $\Delta \Phi$ constant values for the other array plane. In these examples, location of zero phase on the array is chosen to cause a negative beam angle. (Right) The data points correspond to beam angles in the E and H-planes, with the associated beamwidths as errorbars.  These data represent one-quarter of the possible scan positions with $\Delta \Phi_{E/H} = 15$ degrees.}
\end{figure}

The $\Delta \phi$ vs. $\Delta \Phi$ results for the two-dimensional $N \times N = 16 \times 16$ Yagi-Uda array are shown in Figure \ref{fig:phase2d}.  Figure \ref{fig:phase2d} (top left) contains $\Delta \phi_E$ versus $\Delta \Phi_E$ data at 5 GHz.  The data match the theoretical linear slope $\lambda/(2\pi d_y)$ and $\lambda/(2\pi d_z)$, with $d_y = d_z$.  The phase shift per antenna is varied over $[0,75]$ degrees in 15 degree increments independently by row and column.  The circles and squares correspond to $\Delta \Phi_H = 0$ and $45$ degrees, respectively.  Both the circles and squares follow the same line, implying the correct phase independence: when $\Delta \Phi_H$ is held at either constant, $\Delta \phi_E$ still varies with $\Delta \Phi_E$ correctly.  Figure \ref{fig:phase2d} (bottom left) contains $\Delta \phi_H$ versus $\Delta \Phi_H$ data at 5 GHz.  The circles and squares correspond to $\Delta \Phi_E = 0$ and $45$ degrees, respectively.  Beyond $\Delta \Phi_E = 75$ degrees or $\Delta \Phi_H = 75$ degrees, side lobes appear ($>-15$ dB).  Figure \ref{fig:phase2d} (right) contains the beam angle results after steering the beam to 36 of the possible $11 \times 11$ positions in the E and H plane using increments of $\Delta \Phi_{E/H} = 15$ degrees.  The xy-errorbars correspond to the beamwidths.

\begin{figure}
\centering
\includegraphics[width=0.49\textwidth]{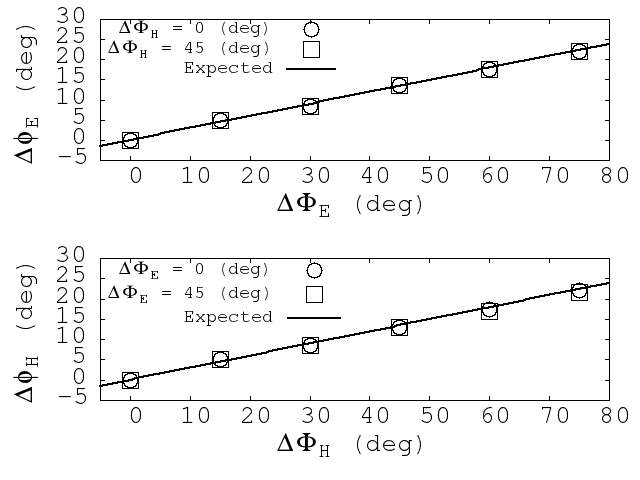}
\includegraphics[width=0.49\textwidth]{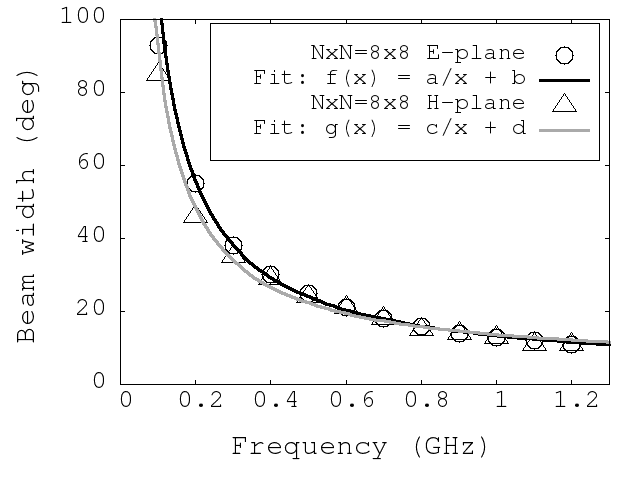}
\caption{\label{fig:phase2d2} (Left) The beam angles $\Delta \phi_E$ and $\Delta \phi_H$ versus the phase shifts per column (for the E-plane) and per row (for the H-plane) for the $N \times N = 8 \times 8$ horn array at 1 GHz.  The black lines represent the theoretical prediction.  In each graph, the circles and squares correspond to two different $\Delta \Phi$ constant values for the other array plane. (Right) The beamwidth in the E and H-planes versus frequency.}
\end{figure}

The $\Delta \phi$ vs. $\Delta \Phi$ results for the two-dimensional $N \times N = 8 \times 8$ horn array are shown in Figure \ref{fig:phase2d2} (left).  Figure \ref{fig:phase2d2} (top left) contains $\Delta \phi_E$ versus $\Delta \Phi_E$ data at 1 GHz.  The larger horn size relative to those in Section \ref{sec:1d} means the upper frequency is $\approx 1.2$ GHz. The data match the theoretical slopes just as in Figure \ref{fig:phase2d}.  The phase shift per antenna is varied in the same pattern as in Figure \ref{fig:phase2d}.  The circles and squares correspond to $\Delta \Phi_H = 0$ and $45$ degrees, respectively, and both data sets follow the theory.  Figure \ref{fig:phase2d2} (bottom left) contains $\Delta \phi_H$ versus $\Delta \Phi_H$ data at 1 GHz.  The circles and squares correspond to $\Delta \Phi_E = 0$ and $45$ degrees, respectively, and both data sets follow the theory.  The beamwidth as a function of frequency across the bandwidth for the design is shown in Figure \ref{fig:phase2d2} (right).  The fit parameter mean values and standard errors are: $a = 10.6 \pm 0.2$ degree GHz, $b = 2.8 \pm 0.3$ degrees, $c = 8.7 \pm 0.4$ degree GHz, and $d = 5.0 \pm 0.9$ degrees.  The width of the mouth of the horns is 16 cm in the E-plane direction and 2 cm in the H-plane direction, so some small difference in beamwidth is not surprising.

\subsection{Radiation Patterns}
\label{sec:2d_rp}

\begin{figure}
\centering
\includegraphics[width=0.3\textwidth]{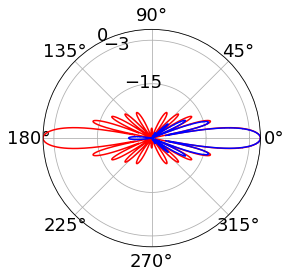}
\includegraphics[width=0.3\textwidth]{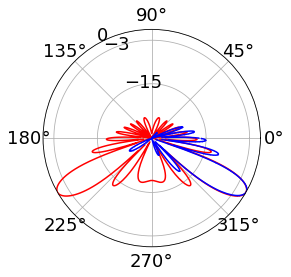}
\includegraphics[width=0.3\textwidth]{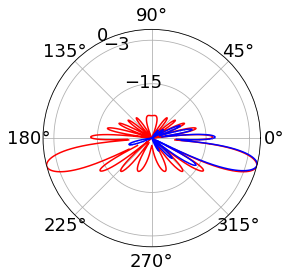} \\
\includegraphics[width=0.3\textwidth]{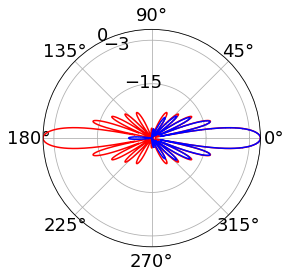}
\includegraphics[width=0.3\textwidth]{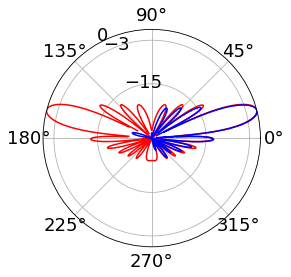}
\includegraphics[width=0.3\textwidth]{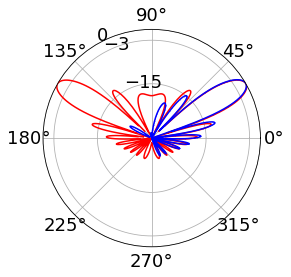} \\
\includegraphics[width=0.3\textwidth]{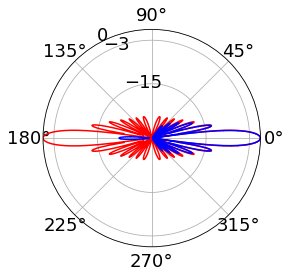}
\includegraphics[width=0.3\textwidth]{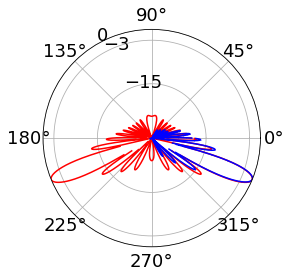}
\includegraphics[width=0.3\textwidth]{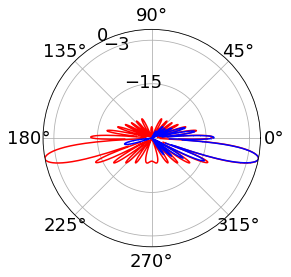} \\ 
\includegraphics[width=0.3\textwidth]{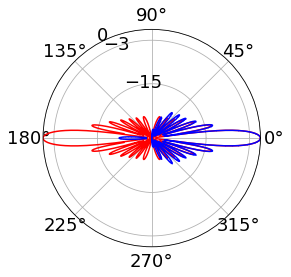}
\includegraphics[width=0.3\textwidth]{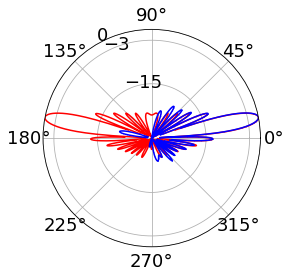}
\includegraphics[width=0.3\textwidth]{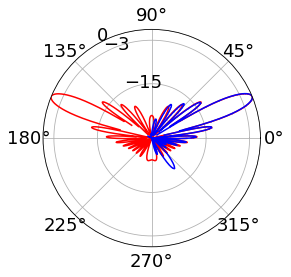} \\
\caption{\label{fig:2dyagipatt} \textbf{Yagi-Uda results, two-dimensional array}(First row) $f = 3$ GHz, with (from left to right) E-plane $(\Delta \Phi_E,\Delta \Phi_H) = (0,0)$ degrees, E-plane $(\Delta \Phi_E,\Delta \Phi_H) = (60,30)$ degrees, E-plane $(\Delta \Phi_E,\Delta \Phi_H) = (30,60)$ degrees. (Second row) $f = 3$ GHz, with (from left to right) H-plane $(\Delta \Phi_E,\Delta \Phi_H) = (0,0)$ degrees, H-plane $(\Delta \Phi_E,\Delta \Phi_H) = (60,30)$ degrees, H-plane $(\Delta \Phi_E,\Delta \Phi_H) = (30,60)$ degrees.  (Third row) $f = 4$ GHz, with (from left to right) E-plane $(\Delta \Phi_E,\Delta \Phi_H) = (0,0)$ degrees, E-plane $(\Delta \Phi_E,\Delta \Phi_H) = (60,30)$ degrees, E-plane $(\Delta \Phi_E,\Delta \Phi_H) = (30,60)$ degrees. (Fourth row) $f = 4$ GHz, with (from left to right) H-plane $(\Delta \Phi_E,\Delta \Phi_H) = (0,0)$ degrees, H-plane $(\Delta \Phi_E,\Delta \Phi_H) = (60,30)$ degrees, H-plane $(\Delta \Phi_E,\Delta \Phi_H) = (30,60)$ degrees. }
\end{figure}

The radiation patterns in the E and H plane for the two-dimensional Yagi-Uda array are shown in Figure \ref{fig:2dyagipatt}.  The phase combinations $(\Delta \Phi_E,\Delta \Phi_H) = (0,0), (30,60), (60,30)$ degrees are shown for E and H planes at 3 and 4 GHz.  As in Section \ref{sec:1d_rp}, Equation \ref{eq:radpatt} is shown in red, and the simulation results are shown in blue.  The main beam and first several side lobes are modeled correctly in each case, and the FB ratio is $\leq -15$ dB.  The side lobes are also at the $\approx -15$ dB level.  Following Figure \ref{fig:phase2d2} (right), the main beam is narrower at 4 GHz than at 3 GHz. Though not generally designed to be broadband elements, the Yagi-Uda elements do display some flexibility in frequency.  The log-periodic dipole array (LPDA) is a broadband example constructed from dipoles as the Yagi is \cite{10.1016/j.astropartphys.2014.09.002}.

\begin{figure}
\centering
\includegraphics[width=0.3\textwidth]{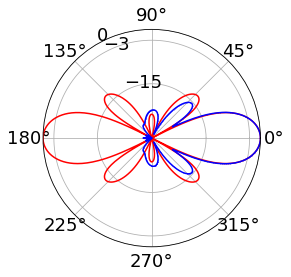}
\includegraphics[width=0.3\textwidth]{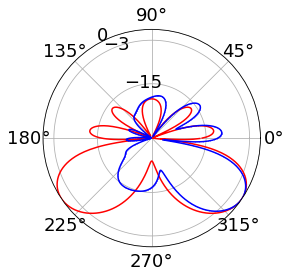}
\includegraphics[width=0.3\textwidth]{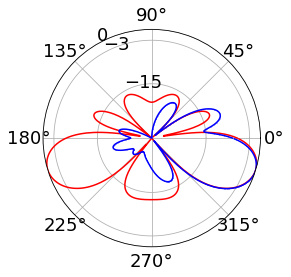} \\
\includegraphics[width=0.3\textwidth]{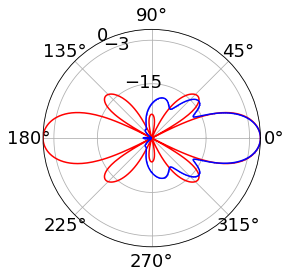}
\includegraphics[width=0.3\textwidth]{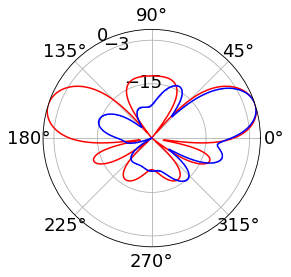}
\includegraphics[width=0.3\textwidth]{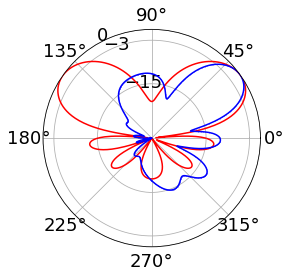} \\
\includegraphics[width=0.3\textwidth]{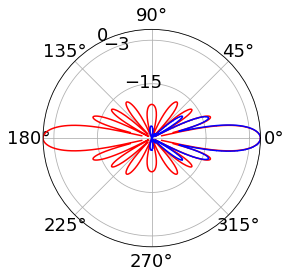}
\includegraphics[width=0.3\textwidth]{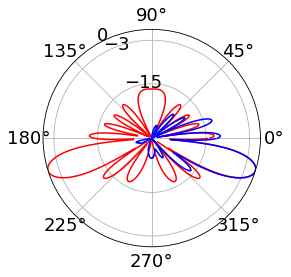}
\includegraphics[width=0.3\textwidth]{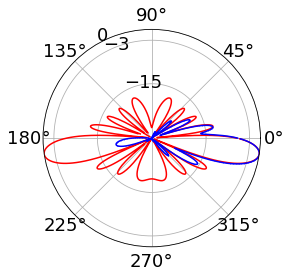} \\
\includegraphics[width=0.3\textwidth]{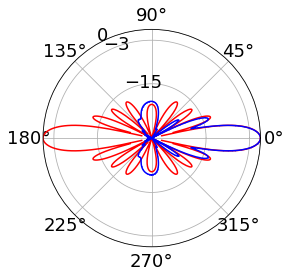}
\includegraphics[width=0.3\textwidth]{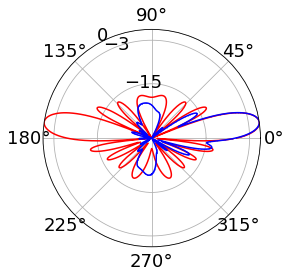}
\includegraphics[width=0.3\textwidth]{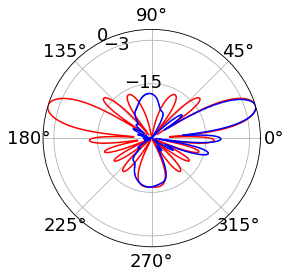} \\
\caption{\label{fig:2dhornpatt} \textbf{Horn results, two-dimensional array} (First row) $f = 0.5$ GHz, with (from left to right) E-plane $(\Delta \Phi_E,\Delta \Phi_H) = (0,0)$ degrees, E-plane $(\Delta \Phi_E,\Delta \Phi_H) = (60,30)$ degrees, E-plane $(\Delta \Phi_E,\Delta \Phi_H) = (30,60)$ degrees. (Second row) $f = 0.5$ GHz, with (from left to right) H-plane $(\Delta \Phi_E,\Delta \Phi_H) = (0,0)$ degrees, H-plane $(\Delta \Phi_E,\Delta \Phi_H) = (60,30)$ degrees, H-plane $(\Delta \Phi_E,\Delta \Phi_H) = (30,60)$ degrees.  (Third row) $f = 1.0$ GHz, with (from left to right) E-plane $(\Delta \Phi_E,\Delta \Phi_H) = (0,0)$ degrees, E-plane $(\Delta \Phi_E,\Delta \Phi_H) = (60,30)$ degrees, E-plane $(\Delta \Phi_E,\Delta \Phi_H) = (30,60)$ degrees. (Fourth row) $f = 1.0$ GHz, with (from left to right) H-plane $(\Delta \Phi_E,\Delta \Phi_H) = (0,0)$ degrees, H-plane $(\Delta \Phi_E,\Delta \Phi_H) = (60,30)$ degrees, H-plane $(\Delta \Phi_E,\Delta \Phi_H) = (30,60)$ degrees. }
\end{figure}

Producing the radiation patterns in Figure \ref{fig:2dyagipatt} requires only $\mathcal{O}(10)$ seconds to run near-field calculations, and only another $\approx 5$ minutes each to run the \verb+get_farfield+ routine over the E and H-planes.  Modeling arrays constructed from dipole elements is orders of magnitude faster than for the array of horns, due to the two-dimensional nature of the dipole elements.  Producing the radiation patterns of Figure \ref{fig:2dhornpatt} for the two-dimensional horn array requires $\approx 60$ minutes combined for the E and H-plane patterns, \textit{per frequency}.  Unlike the Yagi case, the vast majority of time is not dedicated to the \verb+get_farfield+ routine, but to the near-field calculations.  The near-field calculations require ``sub-pixel smoothing'' for the many edges of the blocks that comprise the horn structure.

The radiation patterns in the E and H plane for the two-dimensional horn array are shown in Figure \ref{fig:2dhornpatt}.  The phase combinations $(\Delta \Phi_E,\Phi_H) = (0,0), (30,60), (60,30)$ degrees are shown for E and H planes at 0.5 and 1.0 GHz.  As in Section \ref{sec:1d_rp}, Equation \ref{eq:radpatt} is shown in red, and the simulation results are shown in blue.  The main beam and first several side lobes are modeled correctly in each case, and the FB ratio is $\leq -15$ dB.  The side lobes are also at the $\approx -15$ dB level.  Due to the higher bandwidth, a wider range of beamwidths is available (see Figure \ref{fig:phase2d2}).  The main beam is narrower at 1 GHz than at 0.5 GHz. The horns produce the correct pattern for $(\Delta \Phi_E, \Delta \Phi_H) = (0,0)$ degrees from 0.1 to 1.2 GHz.  However, grating lobes above $-15$ dB are a known problem that occur when attempting to steer phased arrays built from broadband horns to wide angles (see Chapter 9 of Reference \cite{mailloux3}).  The addition of the backplane limits diffraction of the radiation around the edges of the array and therefore limits the FB ratio, but grating lobes appear at $\pm 45$ degrees from the main beam.  There is occasionally a back lobe, which can be attributed to the diffraction of fields around the edge of the backplane.  This effect is more pronounced when the main beam is steered to a wide angle and occurs in the hemisphere opposite to the main beam.

\section{Variation of the Index of Refraction}
\label{sec:n}

The behavior of a one-dimensional phased array embedded within a dielectric medium with spatially-dependent index of refraction $n(z)$ is interesting to the ultra-high energy (UHE) neutrino community \cite{10.1088/1475-7516/2016/02/005,avva}.  Phased arrays represent an opportunity to lower the RF detection threshold for RF pulses generated by UHE neutrinos via the Askaryan effect.  Antarctic ice is the most convenient and natural medium for Askaryan pulse detection, due to the RF transparency and large pristine volumes located in Antarctic and Greenlandic ice sheets and shelves \cite{10.7529/ICRC2011/V04/0340,10.3189/2015jog14j214,10.3189/2015jog15j057}.  The index of refraction varies within the ice because of the transition between surface snow ($\rho \approx 0.4$ g/cm$^3$) and the solid ice below ($\rho = 0.917$ g/cm$^3$).  Most recent and intricate studies of phased array beam behavior still assume a uniform medium \cite{10.1002/mop.32231,10.1038/s41598-019-54120-2,10.1109/array.2016.7832612}.  Embedded phased arrays with varying $n(z)$ emit signals that curve in the direction of increasing $n(z)$.

The \textit{shadow zone} is the volume of ice from which RF signals do not reach a receiver due to the excess curvature of the ray trace \cite{kamlesh}.  While there is evidence that RF signals can propagate horizontally through Antarctic ice \cite{Barwick:2018497}, data from Greenland suggests the relative strength of the effect is small compared to the curved radiation \cite{cosmin}.  Using the tools developed in this work, it is possible to map out the shadow zone for an embedded phased array radiating sinusoidal signals at fixed frequency.  Intriguingly, when the phased array \textit{radiates}, the grating lobe power reflect downward from the snow-air interface, and radiates into the shadow zone.  Grating lobe power also refracts into the air above the interface.  Grating lobe power leaves the array at a different angle than the main beam, so their presence in the shadow zone does not represent forbidden RF propagation.

A two-parameter fit to the $n(z)$ data versus depth $z$ below the surface is given by \cite{Barwick:2018497}

\begin{equation}
  n(z) =
  \begin{cases}
	1 & z > 0 \\
	n_{\rm ice} - \Delta n \exp(z/z_0) & z \leq 0
  \end{cases} \label{eq:n}
\end{equation}

The fit parameters in Equation \ref{eq:n} come from Reference \cite{Barwick:2018497}: $\Delta n = 0.423 \pm 0.004$ and $z_0 = 77 \pm 2$ meters, with $n_{\rm ice} = 1.78$ for RF frequencies.  These values are derived from the SPICE core data taken in 2015 near the South Pole, and are in statistical agreement with fits from data obtained by the RICE experiment (see also Reference \cite{Barwick:2018497}).  Equation \ref{eq:n} was implemented in the one-dimensional horn array case, but the horn structure surrounding each radiating element was removed.  The array is therefore a one-dimensional dipole array.  Further, the length scale was reinterpreted to be meters rather than centimeters, which is an ability conferred by the scale invariant FDTD algorithms.  In this medium, there is no fixed \textit{in-situ} value of $\lambda$ so the $\lambda/4$ dipoles were spaced by $\lambda/2$ according to their free space $\lambda$ value.  At the selected frequency of 200 MHz, the dipole length is 0.375 meters, and the spacing is 0.75 meters.

\begin{figure}[ht]
\centering
\includegraphics[width=10cm,height=4cm]{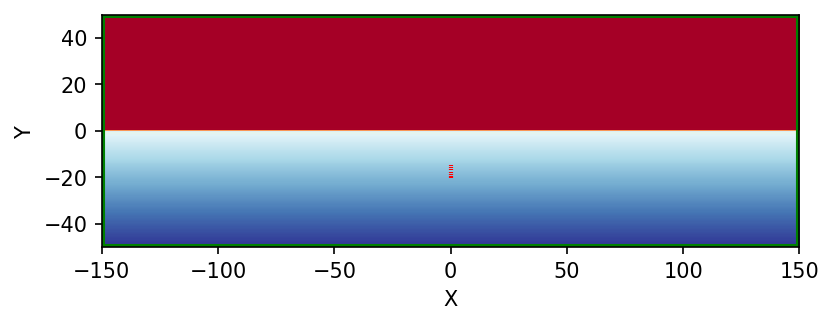}
\caption{\label{fig:deep1} The simplified $N = 8$ one-dimensional vertical phased array with dimensions in meters.  For this array, $d_z = \lambda/2.0$, and the length of the dipole radiators is $\lambda/4.0$.  The colorscale represents $n(z)$ in Equation \ref{eq:n}.}
\end{figure}

\begin{figure}[hb]
\centering
\includegraphics[width=7.5cm,height=2.5cm]{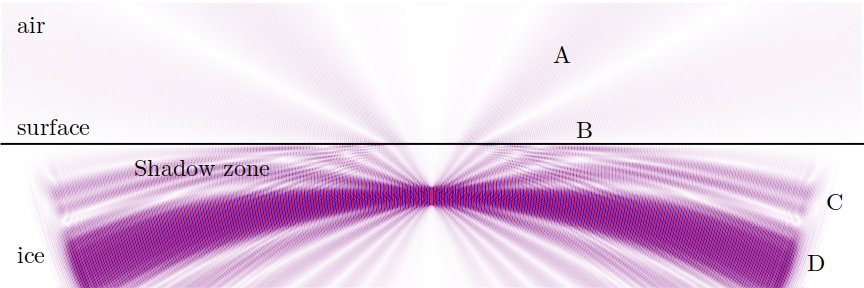}
\includegraphics[width=7.5cm,height=2.5cm]{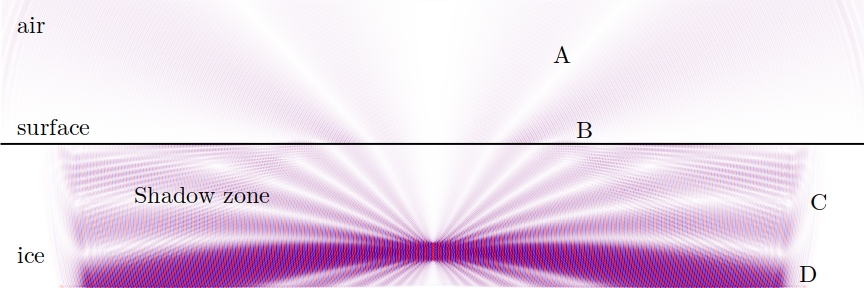}
\caption{\label{fig:deep2} The magnitude of the z-component of the z-polarized dipoles as they radiate as a phased array with $\Delta \Phi = 0$ degrees. The air, surface, and ice regions are the same as Figure \ref{fig:deep1}, with the same dimensions. (Top) The array depth is -15 meters.  (Bottom) The array depth is -35 meters.  (A) Radiation refracting through the surface into the air. (B) Grating lobes reflect from the surface into the shadow zone. (C) Grating lobes propagating through the shadow zone. (D) The main beam bent downward due to the gradient in $n(z)$.}
\end{figure}

Figures \ref{fig:deep1} and \ref{fig:deep2} contain the results of a $N = 8$ one-dimensional dipole array embedded in a medium with the index profile in Equation \ref{eq:n}.  Figure \ref{fig:deep1} shows the schematic of the calculation, and Figure \ref{fig:deep2} shows the magnitude of the z-component of the z-polarized array.  Figure \ref{fig:deep2} represents the same physical dimensions as Figure \ref{fig:deep1}.  Equation \ref{eq:n} was sampled 100 times vertically, and with a resolution parameter of 10, the effective $\Delta z$ is 0.1 meters.  The units in Figure \ref{fig:deep1} are meters, and the unit-less frequency in MEEP was scaled accordingly, to correspond to 200 MHz.  The distances between the air-snow interface and the first phased array element is 15 meters (top) and 35 meters (bottom).  

The color scale in Figure \ref{fig:deep2} is $\pm 0.05$ with the signal amplitude of the elements $\pm 1.0$ at 200 MHz.  The amplitude scale is less important than observing \textit{where} the radiation has penetrated the ice after 200 time steps.  In Figure \ref{fig:deep2} (top), the main beam has curved downwards in the direction of increasing $n(z)$, while grating lobes have both diffracted to the air and reflected into the shadow zone.  The rate of curvature of the main beam is controlled by the fit parameter $z_0$ in Equation \ref{eq:n}.  In Figure \ref{fig:deep2} (bottom), the physics is the same as Figure \ref{fig:deep2} (top), but the effect of $n(z)$ curvature is weakened.  The beam travels farther horizontally because the gradient of $n(z)$ is smaller at the larger depth.  The geometry of the larger depth is such that the reflected grating lobe power is interfering with grating lobe power that was curved downwards without reflection.  This can be seen just above the $C$ marker in Figure \ref{fig:deep2} (bottom).

\section{Summary and Future Analysis}
\label{sec:summary}

Four phased array designs have been modeled with the MIT Electromagnetics Equation Propagation (MEEP) package in non-parallel mode.  Two types of individual radiating element were explored: the narrow-band Yagi-Uda and broadband horn antennas.  Two phased array geometries were explored: \textit{one-dimensional} and \textit{two-dimensional}.  The one and two-dimensional Yagi-Uda phased arrays were designed for $\leq 5$ GHz, however, scale-invariance makes this design scalable to a variety of frequencies.  The one-dimensional horn array performs in the range [0.3 - 5] GHz, using two-dimensional versions of the horn elements.  The two-dimensional horn array had to be modified due to memory constraints.  The result was an array that performed in the range [0.1 - 1.2] GHz.  In all cases, comparisons to array theory were shown.

The one-dimensional array of Yagi-Uda antennas was analyzed in Section \ref{sec:1d}.  The array demonstrated the correct linear relationship between $\Delta \phi$ and $\Delta \Phi$ (Figure \ref{fig:dphase1}) (top left and right).  Although any row of point sources would obey the relationship in Equation \ref{eq:lin}, a row of point sources has two main beam solutions by symmetry.  Thus Figure \ref{fig:dphase1} could not be interpreted correctly were it not for the proper functioning of the Yagi elements.  The radiation patterns produced with the one-dimensional Yagi array were compared to Equation \ref{eq:radpatt} in Figure \ref{fig:1dyagiresults}.  The radiation pattern in the E-plane is shown to agree with Equation \ref{eq:radpatt} in both the main beam and the first several grating lobes.  The calculation takes place in two-dimensions, so an H-plane comparison is not relevant.

The one-dimensional array of horn antennas was analyzed in Section \ref{sec:1d}.  The array demonstrated the correct linear relationship between $\Delta \phi$ and $\Delta \Phi$ (Figure \ref{fig:dphase1}).  In that case, the slope of $\Delta \phi$ vs. $\Delta \Phi$ was \textit{increased} by a factor of 2 and then 10 by \textit{decreasing} the frequency by a factor of 2 and then 10.  The bandwidth of the two-dimensional versions of the horns allows the variation of scan range.  The scan range is smaller at high frequencies, as indicated in Figure \ref{fig:dphase1} (bottom left).  However, the beamwidth is also smaller at high frequencies, as indicated in Figure \ref{fig:dphase1} (bottom right).  The design trade-off is between small beamwidth and large scan range.  In Figure \ref{fig:1dhornresults} the one-dimensional horn array radiation pattern is shown to match Equation \ref{eq:radpatt} at both 0.5 and 5.0 GHz.  There are 2-4 side lobes at 0.5 GHz to match per pattern, and the simulation results match them as well as the wide main beam.  At 5.0 GHz, the main beam is accompanied by two prominent grating lobes at $\pm 45$ degrees that should be as powerful as the main beam.  The simulation finds them at the -15 dB level.  The grating lobes are being suppressed by the the pattern null from the horn element pattern \cite{mailloux3}.  At lower frequencies, however, scan loss takes a toll on radiated power (Tab. \ref{tab:runParam}).

The two-dimensional, $N \times N = 16 \times 16$ Yagi-Uda array was analyzed in Section \ref{sec:2d}.  The array demonstrated the correct linear relationships between $\Delta \phi_E$ and $\Delta \Phi_E$, and $\Delta \phi_H$ and $\Delta \Phi_H$ (Figure \ref{fig:phase2d}) (left).  Given the narrow beamwidth, the array design can be scanned $\pm 5$ beamwidths in the E-plane and $\pm 5$ beamwidths in the H-plane before side lobes become too large.  One fourth of these scan positions are shown in Figure \ref{fig:phase2d} (right).  The radiation pattern of the full two-dimensional array was displayed in Figure \ref{fig:2dyagipatt} at 3 and 4 GHz, for several scan angles.  In each case, the pattern matched Equation \ref{eq:radpatt} in the E and H-planes in the main beam and dominant side lobes.  The addition of a metal back plane helps to suppress back lobes.  Peculiarly, the two-dimensional array did not match the theoretical prediction at 5 GHz as well as the one-dimensional case when scanned.

The two-dimensional, $N \times N = 8 \times 8$ horn array was analyzed in Section \ref{sec:2d}.  The array demonstrated the correct linear relationships between $\Delta \phi_E$ and $\Delta \Phi_E$, and $\Delta \phi_H$ and $\Delta \Phi_H$ (Figure \ref{fig:phase2d2}) (left).  The beamwidth is again inversely proportional to frequency (Figure \ref{fig:phase2d2}) (right).  It is not surprising that the fits differ slightly in the E and H-planes, since the horn width changes in the E-plane but does not in the H-plane.  The quality of the fits to $1/f + const$ are excellent.  The additive constants in these fits are only necessary because the array cannot be infinitely long.  Technically, Equation \ref{eq:radpatt} implies that the beamwidth would go to zero as $N \rightarrow \infty$.  The radiation patterns of the two-dimensional horn array are displayed in Figure \ref{fig:2dhornpatt} at $0.5$ and $1.0$ GHz, for the same sampling of scan angles as in Figure \ref{fig:2dyagipatt}.  The high-frequency beam is narrower and accompanied by grating lobes at $\pm 45$ degrees.  The patterns agree with theoretical expectations, with the exception of the H-plane lobes at $\pm 90$ degrees.  At low frequency, the beam is wider and is accompanied by grating lobes at $\pm 45$ degrees from the main beam.  The results match in the main lobe, but the simulation does not match the theoretical grating lobes.  This is pronounced when the beam is moved far from broadside in the H-plane.

Finally, a simplified version of the $N = 8$ one-dimensional case of dipoles was embedded in a medium with varying index of refraction, $n(z)$.  The model for $n(z)$ was a simple fit to the profile of the ice at the South Pole, which is a location of interest for planned phased array detectors designed to record Askaryan signals from UHE neutrinos passing through ice.  Though the studies in this work are restricted to phased-arrays as transmitters, and not receivers, the shadow zone of the array was mapped at 200 MHz under realistic conditions.  An interesting side effect of the phased array being the radiating system was that the grating lobes managed to propagate into the shadow zone.

Future work would include several enhancements to the simulations.  Calculations of S-parameters for individual elements should be added, and optimization studies on horn and Yagi geometric parameters are warranted.  However, other RF element types should also be studied.  Due to the relevance of one-dimensional phased array receivers for UHE neutrino physics, one interesting choice is the wide-radius dipole used by the Radio Neutrino Observatory Greenland (RNO-G) collaboration \cite{rnog}.  Such elements already have low VSWR measurements in the relevant bandwidth.  Finally, upgrading the simulation code to utilize parallel MEEP capabilities will increase the potential speed and complexity.  Additional complexity will come in the form of more accurate antenna structure modeling, thereby improving the trustworthiness across a wide bandwidth.

\vspace{6pt} 

\funding{This research was funded by the Office of Naval Research (ONR) under the Summer Faculty Research Program (SFRP).}

\institutionalreview{Not applicable.}
\informedconsent{Not applicable.}
\dataavailability{The data presented in this study are available on request from the corresponding author. Due to large file sizes and restricted bandwidth, please contact corresponding author to set up collaborative sharing.}

\acknowledgments{We would like to thank the Office of Naval Research (ONR) for helping to support this research.  In particular, we would like to thank the Naval Surface Warfare Center Corona Division and their continued support of the ONR Summer Faculty Research Program (SFRP).  Conversations with Christopher Clark and Gary Yeakley were especially helpful.  We are also grateful to Van Nguyen, Jeffery Benson, and Golda McWhorter.  Karon Myles deserves our special thanks for helping to coordinate the SFRP program.}

\conflictsofinterest{The authors declare no conflict of interest.}

\appendixtitles{no}

\appendix
\section{}
\unskip

MEEP FDTD is more often applied to $1 \mu$m scale lengths than the 1 cm-scale RF elements, so scale-invariance must be highlighted.  Scale invariant units with $c = 1$ are used in MEEP when solving Maxwell's equations with the FDTD technique.  The typical length scale in MEEP analysis is usually called the ``a-value.''  Systems with dimensions of order $1 \mu$m are said to have an ``a-value of $1 \mu$m.''  A value of $a = 1$ cm is chosen for models presented in Secs. \ref{sec:1d} and \ref{sec:2d}.  For example, if the frequency $f = 7.5$ GHz, the wavelength is $4.0$ cm $= 4.0a$, or simply $4.0$ with $f=1/\lambda = 0.25$.  Since $c=1$, $\lambda = f^{-1} = T$.  The period is $4.0$, so a simulation run of $50 T = 200$ time units corresponds to 6.67 ns.  Assuming 1 pixel/a-value, simulated radiation would therefore propagate 200 units of $\Delta x$ in a straight line before time was up.  A \textit{resolution} parameter sets the number of pixels per distance unit and is usually larger than 1.0.  Selecting the right resolution is often a subtle balance between capturing the most relevant effects while limiting the memory usage of the simulation results.

\reftitle{References}

\externalbibliography{yes}
\bibliography{mybibfile.bib}

\begin{thebibliography}{-------}
\providecommand{\natexlab}[1]{#1}

\bibitem[Syrytsin \em{et~al.}(2017)Syrytsin, Zhang, and
  Pedersen]{10.1109/iceaa.2017.8065458}
Syrytsin, I.; Zhang, S.; Pedersen, G.F.
\newblock {Circularly Polarized Planar Helix Phased Antenna Array for 5G Mobile
  Terminals}.
\newblock {\em 2017 International Conference on Electromagnetics in Advanced
  Applications (ICEAA), Verona, Italy, September 2017} {\bf 2017}, pp.
  1105--1108.
\newblock
  doi:{\changeurlcolor{black}\href{https://doi.org/10.1109/iceaa.2017.8065458}{\detokenize{10.1109/iceaa.2017.8065458}}}.

\bibitem[Kikuchi \em{et~al.}(2017)Kikuchi, Mikada, and
  Takekawa]{10.3997/2214-4609.201701121}
Kikuchi, K.; Mikada, H.; Takekawa, J.
\newblock {Improved Imaging Capability of Phased Array Antenna in Ground
  Penetrating Radar Survey}.
\newblock {\em Conference Proceedings, 79th EAGE Conference and Exhibition
  2017, Paris, France, June 2017} {\bf 2017}.
\newblock
  doi:{\changeurlcolor{black}\href{https://doi.org/10.3997/2214-4609.201701121}{\detokenize{10.3997/2214-4609.201701121}}}.

\bibitem[Vieregg \em{et~al.}(2016)Vieregg, Bechtol, and
  Romero-Wolf]{10.1088/1475-7516/2016/02/005}
Vieregg, A.; Bechtol, K.; Romero-Wolf, A.
\newblock A technique for detection of {PeV} neutrinos using a phased radio
  array.
\newblock {\em Journal of Cosmology and Astroparticle Physics} {\bf 2016}, {\em
  2016},~005.
\newblock
  doi:{\changeurlcolor{black}\href{https://doi.org/10.1088/1475-7516/2016/02/005}{\detokenize{10.1088/1475-7516/2016/02/005}}}.

\bibitem[Munekata \em{et~al.}(2014)Munekata, Yamamoto, and
  Nojima]{10.1109/iwem.2014.6963645}
Munekata, T.; Yamamoto, M.; Nojima, T.
\newblock {A Wideband 16-Element Antenna Array Using Leaf-Shaped Bowtie Antenna
  and Series-Parallel Feed Networks}.
\newblock {\em 2014 IEEE International Workshop on Electromagnetics (iWEM),
  Sapporo Hokkaido, Japan, August 2014} {\bf 2014}, pp. 80--81.
\newblock
  doi:{\changeurlcolor{black}\href{https://doi.org/10.1109/iwem.2014.6963645}{\detokenize{10.1109/iwem.2014.6963645}}}.

\bibitem[Avva \em{et~al.}(2017)Avva, Bechtol, Chesebro, Cremonesi, Deaconu,
  Gupta, Ludwig, Messino, Miki, Nichol, Oberla, Ransom, Romero-Wolf, Saltzberg,
  Schlupf, Shipp, Varner, Vieregg, and Wissel]{avva}
Avva, J.; Bechtol, K.; Chesebro, T.; Cremonesi, L.; Deaconu, C.; Gupta, A.;
  Ludwig, A.; Messino, W.; Miki, C.; Nichol, R.; Oberla, E.; Ransom, M.;
  Romero-Wolf, A.; Saltzberg, D.; Schlupf, C.; Shipp, N.; Varner, G.; Vieregg,
  A.; Wissel, S.
\newblock Development toward a ground-based interferometric phased array for
  radio detection of high energy neutrinos.
\newblock {\em Nuclear Instruments and Methods in Physics Research Section A:
  Accelerators, Spectrometers, Detectors and Associated Equipment} {\bf 2017},
  {\em 869},~46 -- 55.
\newblock
  doi:{\changeurlcolor{black}\href{https://doi.org/https://doi.org/10.1016/j.nima.2017.07.009}{\detokenize{https://doi.org/10.1016/j.nima.2017.07.009}}}.

\bibitem[{Ansys, Inc., Canonsburg, Pennsylvania}(2020)]{hfss}
{Ansys, Inc., Canonsburg, Pennsylvania}.
\newblock {\em 3D Electromagnetic Field Simulator for RF and Wireless Design},
  2020.

\bibitem[Feng \em{et~al.}(2019)Feng, Zhang, Tian, Zhu, Joines, and
  Wang]{10.1109/tmtt.2019.2919838}
Feng, N.; Zhang, Y.; Tian, X.; Zhu, J.; Joines, W.T.; Wang, G.P.
\newblock {System-Combined ADI-FDTD Method and Its Electromagnetic Applications
  in Microwave Circuits and Antennas}.
\newblock {\em IEEE Transactions on Microwave Theory and Techniques} {\bf
  2019}, {\em 67},~3260--3270.
\newblock
  doi:{\changeurlcolor{black}\href{https://doi.org/10.1109/tmtt.2019.2919838}{\detokenize{10.1109/tmtt.2019.2919838}}}.

\bibitem[Zhu \em{et~al.}(2017)Zhu, Hwang, Ren, and
  Yang]{10.1109/apusncursinrsm.2017.8072977}
Zhu, L.; Hwang, H.S.; Ren, E.; Yang, G.
\newblock {High Performance MIMO Antenna for 5G Wearable Devices}.
\newblock {\em 2017 IEEE International Symposium on Antennas and Propagation \&
  USNC/URSI National Radio Science Meeting, San Diego, CA, USA, July 2017} {\bf
  2017}, pp. 1869--1870.
\newblock
  doi:{\changeurlcolor{black}\href{https://doi.org/10.1109/apusncursinrsm.2017.8072977}{\detokenize{10.1109/apusncursinrsm.2017.8072977}}}.

\bibitem[Ho \em{et~al.}()Ho, Hunt, Hewett, Ready, Mittra, Yu, Zolnick, and
  Kragalott]{10.1109/aps.2006.1710856}
Ho, T.Q.; Hunt, L.N.; Hewett, C.A.; Ready, T.G.; Mittra, R.; Yu, W.; Zolnick,
  D.A.; Kragalott, M.
\newblock {Analysis of Electrically Large Patch Phased Arrays via CFDTD}.
\newblock {\em 2006 IEEE Antennas and Propagation Society International
  Symposium, Albuquerque, New Mexico, July 2006}.

\bibitem[Burke \em{et~al.}(2004)Burke, Miller, and
  Poggio]{10.1109/aps.2004.1331976}
Burke, G.J.; Miller, E.K.; Poggio, A.J.
\newblock {The Numerical Electromagnetics Code (NEC) - A Brief History}.
\newblock {\em IEEE Antennas and Propagation Society Symposium, 2004} {\bf
  2004}, {\em 3},~2871--2874.
\newblock
  doi:{\changeurlcolor{black}\href{https://doi.org/10.1109/aps.2004.1331976}{\detokenize{10.1109/aps.2004.1331976}}}.

\bibitem[Oskooi \em{et~al.}(2010)Oskooi, Roundy, Ibanescu, Bermel,
  Joannopoulos, and Johnson]{10.1016/j.cpc.2009.11.008}
Oskooi, A.F.; Roundy, D.; Ibanescu, M.; Bermel, P.; Joannopoulos, J.; Johnson,
  S.G.
\newblock {Meep: A flexible free-software package for electromagnetic
  simulations by the FDTD method}.
\newblock {\em Computer Physics Communications} {\bf 2010}, {\em
  181},~687--702.
\newblock
  doi:{\changeurlcolor{black}\href{https://doi.org/10.1016/j.cpc.2009.11.008}{\detokenize{10.1016/j.cpc.2009.11.008}}}.

\bibitem[Fedeli \em{et~al.}(2019)Fedeli, Montecucco, and
  Gragnani]{10.3390/electronics8121506}
Fedeli, A.; Montecucco, C.; Gragnani, G.L.
\newblock {Open-Source Software for Electromagnetic Scattering Simulation: The
  Case of Antenna Design}.
\newblock {\em Electronics} {\bf 2019}, {\em 8},~1506.
\newblock
  doi:{\changeurlcolor{black}\href{https://doi.org/10.3390/electronics8121506}{\detokenize{10.3390/electronics8121506}}}.

\bibitem[Liebig \em{et~al.}(2013)Liebig, Rennings, Held, and
  Erni]{10.1002/jnm.1875}
Liebig, T.; Rennings, A.; Held, S.; Erni, D.
\newblock {openEMS – a free and open source equivalent‐circuit (EC) FDTD
  simulation platform supporting cylindrical coordinates suitable for the
  analysis of traveling wave MRI applications}.
\newblock {\em International Journal of Numerical Modelling: Electronic
  Networks, Devices and Fields} {\bf 2013}, {\em 26},~680--696.
\newblock
  doi:{\changeurlcolor{black}\href{https://doi.org/10.1002/jnm.1875}{\detokenize{10.1002/jnm.1875}}}.

\bibitem[Warren \em{et~al.}(2016)Warren, Giannopoulos, and
  Giannakis]{10.1016/j.cpc.2016.08.020}
Warren, C.; Giannopoulos, A.; Giannakis, I.
\newblock {gprMax: Open source software to simulate electromagnetic wave
  propagation for Ground Penetrating Radar}.
\newblock {\em Computer Physics Communications} {\bf 2016}, {\em
  209},~163--170.
\newblock
  doi:{\changeurlcolor{black}\href{https://doi.org/10.1016/j.cpc.2016.08.020}{\detokenize{10.1016/j.cpc.2016.08.020}}}.

\bibitem[Richie and Ababei(2017)]{10.1016/j.jcde.2017.06.004}
Richie, J.E.; Ababei, C.
\newblock {Optimization of patch antennas via multithreaded simulated annealing
  based design exploration}.
\newblock {\em Journal of Computational Design and Engineering} {\bf 2017},
  {\em 4},~249--255.
\newblock
  doi:{\changeurlcolor{black}\href{https://doi.org/10.1016/j.jcde.2017.06.004}{\detokenize{10.1016/j.jcde.2017.06.004}}}.

\bibitem[Mailloux(2018)]{mailloux3}
Mailloux, R.
\newblock {\em The Phased Array Antenna Handbook}; Artech House: Norwood, MA,
  2018.

\bibitem[Wahl \em{et~al.}(2013)Wahl, Gagnon, Debaes, Erps, Vermeulen, Miller,
  and Thienpont]{10.2528/pier13030606}
Wahl, P.; Gagnon, D.S.L.; Debaes, C.; Erps, J.V.; Vermeulen, N.; Miller,
  D.A.B.; Thienpont, H.
\newblock {B-CALM: An Open-Source Multi-GPU-based 3D-FDTD with Mutli-pole
  dispersion for Plasmonics}.
\newblock {\em Progress In Electromagnetics Research} {\bf 2013}, {\em
  138},~467--478.
\newblock
  doi:{\changeurlcolor{black}\href{https://doi.org/10.2528/pier13030606}{\detokenize{10.2528/pier13030606}}}.

\bibitem[Gorham \em{et~al.}(2009)Gorham, Allison, Barwick, Beatty, Besson,
  Binns, Chen, Chen, Clem, Connolly, Dowkontt, DuVernois, Field, Goldstein,
  Goodhue, Hast, Hebert, Hoover, Israel, Kowalski, Learned, Liewer, Link,
  Lusczek, Matsuno, Mercurio, Miki, Miočinović, Nam, Naudet, Nichol,
  Palladino, Reil, Romero-Wolf, Rosen, Ruckman, Saltzberg, Seckel, Varner,
  Walz, Wang, Williams, and Wu]{10.1016/j.astropartphys.2009.05.003}
Gorham, P.; Allison, P.; Barwick, S.; Beatty, J.; Besson, D.; Binns, W.; Chen,
  C.; Chen, P.; Clem, J.; Connolly, A.; Dowkontt, P.; DuVernois, M.; Field, R.;
  Goldstein, D.; Goodhue, A.; Hast, C.; Hebert, C.; Hoover, S.; Israel, M.;
  Kowalski, J.; Learned, J.; Liewer, K.; Link, J.; Lusczek, E.; Matsuno, S.;
  Mercurio, B.; Miki, C.; Miočinović, P.; Nam, J.; Naudet, C.; Nichol, R.;
  Palladino, K.; Reil, K.; Romero-Wolf, A.; Rosen, M.; Ruckman, L.; Saltzberg,
  D.; Seckel, D.; Varner, G.; Walz, D.; Wang, Y.; Williams, C.; Wu, F.
\newblock {The Antarctic Impulsive Transient Antenna ultra-high energy neutrino
  detector: Design, performance, and sensitivity for the 2006–2007 balloon
  flight}.
\newblock {\em Astroparticle Physics} {\bf 2009}, {\em 32},~10--41,
  \href{http://xxx.lanl.gov/abs/0812.1920}{{\normalfont [0812.1920]}}.
\newblock
  doi:{\changeurlcolor{black}\href{https://doi.org/10.1016/j.astropartphys.2009.05.003}{\detokenize{10.1016/j.astropartphys.2009.05.003}}}.

\bibitem[Gorham \em{et~al.}(2007)Gorham, Barwick, Beatty, Besson, Binns, Chen,
  Chen, Clem, Connolly, and Dowkontt]{undefined}
Gorham, P.; Barwick, S.; Beatty, J.; Besson, D.; Binns, W.; Chen, C.; Chen, P.;
  Clem, J.; Connolly, A.; Dowkontt, P.
\newblock {Observations of the Askaryan effect in ice}.
\newblock {\em Physical review letters} {\bf 2007}, {\em 99},~171101,
  \href{http://xxx.lanl.gov/abs/hep-ex/0611008}{{\normalfont
  [hep-ex/0611008]}}.

\bibitem[Barwick \em{et~al.}(2015)Barwick, Berg, Besson, Duffin, Hanson, Klein,
  Kleinfelder, Piasecki, Ratzlaff, Reed, Roumi, Stezelberger, Tatar, Walker,
  Young, and Zou]{10.1016/j.astropartphys.2014.09.002}
Barwick, S.; Berg, E.; Besson, D.; Duffin, T.; Hanson, J.; Klein, S.;
  Kleinfelder, S.; Piasecki, M.; Ratzlaff, K.; Reed, C.; Roumi, M.;
  Stezelberger, T.; Tatar, J.; Walker, J.; Young, R.; Zou, L.
\newblock {Time-domain response of the ARIANNA detector}.
\newblock {\em Astroparticle Physics} {\bf 2015}, {\em 62},~139--151,
  \href{http://xxx.lanl.gov/abs/1406.0820}{{\normalfont [1406.0820]}}.
\newblock
  doi:{\changeurlcolor{black}\href{https://doi.org/10.1016/j.astropartphys.2014.09.002}{\detokenize{10.1016/j.astropartphys.2014.09.002}}}.

\bibitem[Hanson(2011)]{10.7529/ICRC2011/V04/0340}
Hanson, J.C.
\newblock {Ross Ice Shelf Thickness, Radio-frequency Attenuation and
  Reflectivity: Implications for the ARIANNA UHE Neutrino Detector}.
\newblock {\em In Proceedings of the 32nd International Cosmic Ray Conference,
  Beijing, China, August 2011} {\bf 2011}.
\newblock
  doi:{\changeurlcolor{black}\href{https://doi.org/10.7529/ICRC2011/V04/0340}{\detokenize{10.7529/ICRC2011/V04/0340}}}.

\bibitem[Hanson \em{et~al.}(2015)Hanson, Barwick, Berg, Besson, Duffin, Klein,
  Kleinfelder, Reed, Roumi, Stezelberger, Tatar, Walker, and
  Zou]{10.3189/2015jog14j214}
Hanson, J.C.; Barwick, S.W.; Berg, E.C.; Besson, D.Z.; Duffin, T.J.; Klein,
  S.R.; Kleinfelder, S.A.; Reed, C.; Roumi, M.; Stezelberger, T.; Tatar, J.;
  Walker, J.A.; Zou, L.
\newblock {Radar absorption, basal reflection, thickness and polarization
  measurements from the Ross Ice Shelf, Antarctica}.
\newblock {\em Journal of Glaciology} {\bf 2015}, {\em 61},~438--446.
\newblock
  doi:{\changeurlcolor{black}\href{https://doi.org/10.3189/2015jog14j214}{\detokenize{10.3189/2015jog14j214}}}.

\bibitem[Avva \em{et~al.}(2014)Avva, Kovac, Miki, Saltzberg, and
  Vieregg]{10.3189/2015jog15j057}
Avva, J.; Kovac, J.; Miki, C.; Saltzberg, D.; Vieregg, A.
\newblock {An in situ measurement of the radio-frequency attenuation in ice at
  Summit Station, Greenland}.
\newblock {\em Journal of Glaciology} {\bf 2014},
  \href{http://xxx.lanl.gov/abs/1409.5413}{{\normalfont [1409.5413]}}.
\newblock
  doi:{\changeurlcolor{black}\href{https://doi.org/10.3189/2015jog15j057}{\detokenize{10.3189/2015jog15j057}}}.

\bibitem[Adamidis \em{et~al.}(2020)Adamidis, Vardiambasis, Ioannidou, and
  Kapetanakis]{10.1002/mop.32231}
Adamidis, G.A.; Vardiambasis, I.O.; Ioannidou, M.P.; Kapetanakis, T.N.
\newblock {Design and implementation of an adaptive beamformer for phased array
  antenna applications}.
\newblock {\em Microwave and Optical Technology Letters} {\bf 2020}, {\em
  62},~1780--1784.
\newblock
  doi:{\changeurlcolor{black}\href{https://doi.org/10.1002/mop.32231}{\detokenize{10.1002/mop.32231}}}.

\bibitem[Ahn \em{et~al.}(2019)Ahn, Hwang, Kim, Chae, Yu, and
  Lee]{10.1038/s41598-019-54120-2}
Ahn, B.; Hwang, I.J.; Kim, K.S.; Chae, S.C.; Yu, J.W.; Lee, H.L.
\newblock {Wide-Angle Scanning Phased Array Antenna using High Gain Pattern
  Reconfigurable Antenna Elements}.
\newblock {\em Scientific Reports} {\bf 2019}, {\em 9},~18391.
\newblock
  doi:{\changeurlcolor{black}\href{https://doi.org/10.1038/s41598-019-54120-2}{\detokenize{10.1038/s41598-019-54120-2}}}.

\bibitem[Gampala and Reddy(2016)]{10.1109/array.2016.7832612}
Gampala, G.; Reddy, C.J.
\newblock {Advanced Computational Tools for Phased Array Antenna Applications}.
\newblock {\em 2016 IEEE International Symposium on Phased Array Systems and
  Technology (PAST), Waltham, MA, USA, October 2016} {\bf 2016}, pp. 1--5.
\newblock
  doi:{\changeurlcolor{black}\href{https://doi.org/10.1109/array.2016.7832612}{\detokenize{10.1109/array.2016.7832612}}}.

\bibitem[Dookayka(2011)]{kamlesh}
Dookayka, K.
\newblock Characterizing the Search for Ultra-High Energy Neutrinos with the
  ARIANNA Detector.
\newblock PhD thesis, Univeristy of California at Irvine,  2011.

\bibitem[Barwick \em{et~al.}(2018)Barwick, Berg, Besson, Gaswint, Glaser,
  Hallgren, Hanson, Klein, Kleinfelder, Köpke, Kravchenko, Lahmann, Latif,
  Nam, Nelles, Persichilli, Sandstrom, Tatar, and Unger]{Barwick:2018497}
Barwick, S.W.; Berg, E.C.; Besson, D.Z.; Gaswint, G.; Glaser, C.; Hallgren, A.;
  Hanson, J.C.; Klein, S.R.; Kleinfelder, S.; Köpke, L.; Kravchenko, I.;
  Lahmann, R.; Latif, U.; Nam, J.; Nelles, A.; Persichilli, C.; Sandstrom, P.;
  Tatar, J.; Unger, E.
\newblock {Observation of classically `forbidden' electromagnetic wave
  propagation and implications for neutrino detection.}
\newblock {\em Journal of Cosmology and Astroparticle Physics} {\bf 2018}, {\em
  2018},~055--055.
\newblock
  doi:{\changeurlcolor{black}\href{https://doi.org/10.1088/1475-7516/2018/07/055}{\detokenize{10.1088/1475-7516/2018/07/055}}}.

\bibitem[Deaconu \em{et~al.}(2018)Deaconu, Vieregg, Wissel, Bowen, Chipman,
  Gupta, Miki, Nichol, and Saltzberg]{cosmin}
Deaconu, C.; Vieregg, A.G.; Wissel, S.A.; Bowen, J.; Chipman, S.; Gupta, A.;
  Miki, C.; Nichol, R.J.; Saltzberg, D.
\newblock Measurements and Modeling of Near-Surface Radio Propagation in
  Glacial Ice and Implications for Neutrino Experiments.
\newblock {\em Phys. Rev. D} {\bf 2018}, {\em 98},~043010.
\newblock
  doi:{\changeurlcolor{black}\href{https://doi.org/10.1103/PhysRevD.98.043010}{\detokenize{10.1103/PhysRevD.98.043010}}}.

\bibitem[{Aguilar} \em{et~al.}(2020){Aguilar}, {Allison}, {Beatty}, {Bernhoff},
  {Besson}, {Bingefors}, {Botner}, {Buitink}, {Carter}, {Clark}, {Connolly},
  {Dasgupta}, {de Kockere}, {de Vries}, {Deaconu}, {DuVernois}, {Feigl},
  {Garcia-Fernandez}, {Glaser}, {Hallgren}, {Hallmann}, {Hanson}, {Hendricks},
  {Hokanson-Fasig}, {Hornhuber}, {Hughes}, {Karle}, {Kelley}, {Klein}, {Krebs},
  {Lahmann}, {Magnuson}, {Meures}, {Meyers}, {Nelles}, {Novikov}, {Oberla},
  {Oeyen}, {Pandya}, {Plaisier}, {Pyras}, {Ryckbosch}, {Scholten}, {Seckel},
  {Smith}, {Southall}, {Torres}, {Toscano}, {Van Den Broeck}, {van Eijndhoven},
  {Vieregg}, {Welling}, {Wissel}, {Young}, and {Zink}]{rnog}
{Aguilar}, J.A.; {Allison}, P.; {Beatty}, J.J.; {Bernhoff}, H.; {Besson}, D.;
  {Bingefors}, N.; {Botner}, O.; {Buitink}, S.; {Carter}, K.; {Clark}, B.A.;
  {Connolly}, A.; {Dasgupta}, P.; {de Kockere}, S.; {de Vries}, K.D.;
  {Deaconu}, C.; {DuVernois}, M.A.; {Feigl}, N.; {Garcia-Fernandez}, D.;
  {Glaser}, C.; {Hallgren}, A.; {Hallmann}, S.; {Hanson}, J.C.; {Hendricks},
  B.; {Hokanson-Fasig}, B.; {Hornhuber}, C.; {Hughes}, K.; {Karle}, A.;
  {Kelley}, J.L.; {Klein}, S.R.; {Krebs}, R.; {Lahmann}, R.; {Magnuson}, M.;
  {Meures}, T.; {Meyers}, Z.S.; {Nelles}, A.; {Novikov}, A.; {Oberla}, E.;
  {Oeyen}, B.; {Pandya}, H.; {Plaisier}, I.; {Pyras}, L.; {Ryckbosch}, D.;
  {Scholten}, O.; {Seckel}, D.; {Smith}, D.; {Southall}, D.; {Torres}, J.;
  {Toscano}, S.; {Van Den Broeck}, D.J.; {van Eijndhoven}, N.; {Vieregg}, A.G.;
  {Welling}, C.; {Wissel}, S.; {Young}, R.; {Zink}, A.
\newblock {Design and Sensitivity of the Radio Neutrino Observatory in
  Greenland (RNO-G)}.
\newblock {\em arXiv e-prints} {\bf 2020}, p. arXiv:2010.12279,
  \href{http://xxx.lanl.gov/abs/2010.12279}{{\normalfont
  [arXiv:astro-ph.IM/2010.12279]}}.

\end{thebibliography}

\end{document}